\documentclass[aps,pre,twocolumn,groupedaddress]{revtex4}
\usepackage{graphicx}
\usepackage{amsmath,amsthm,amssymb}
\usepackage{epsfig}

\begin{document}

\title{Price Trends in a Simplified Model of the Wealth Game}
\author{W. Y. Cheung}
\email{phmaize@ust.hk}
\affiliation{Department of Physics, 
The Hong Kong University of Science and Technology, Hong Kong, China.}
\author{K. Y. Michael Wong}
\email{phkywong@ust.hk}
\affiliation{Department of Physics, 
The Hong Kong University of Science and Technology, Hong Kong, China.}
\date{9 September 2010}

\begin{abstract}
We consider a simplified version of the Wealth Game, 
which is an agent-based financial market model 
with many interesting features resembling the real stock market. 
Market makers are not present in the game 
so that the majority traders are forced to reduce 
the amount of stocks they trade, 
in order to have a balance in the supply and demand. 
The strategy space is also simplified 
so that the market is only left with strategies 
resembling the decisions of optimistic or pessimistic fundamentalists 
and trend-followers in the real stock market. 
A dynamical phase transition between a trendsetters' phase 
and a bouncing phase is discovered 
in the space of price sensitivity and market impact. 
Analysis based on a semi-empirical approach explains the phase transition 
and locates the phase boundary. 
A phase transition is also observed 
when the fraction of trend-following strategies increases, 
which can be explained macroscopically 
by matching the supply and demand of stocks.
\end{abstract}

\maketitle

\section{Introduction}

A recent remarkable trend in the physics community is its engagement
in interdisciplinary fields using physics-inspired techniques. Econophysics
is one such area in which financial markets are simulated by agent-based
models in much the same way as other many-body systems in statistical
physics.
The Minority Game (MG) \cite{Challet97} is an agent-based model based
on the insight that agents making minority decisions in markets can
take advantage of other agents. Due to its success in capturing the
profit-seeking behavior of agents, it became the progenitor of a family
of agent-based models \cite{Challet05}, which study various aspects
of market behavior, such as volatility \cite{Savit99}, noise \cite{Cavagna99},
market-clearing mechanisms \cite{Jefferies01}, and anticipative strategies
\cite{Andersen03}.

The market behavior depends on the way 
the agents evaluate their strategies 
when they make choices among them to take actions. 
In early versions of the minority games, agents evaluate their strategies
using various {\it virtual points} or {\it scores}. Typical virtual
point updating rules, such as those in the original MG \cite{Challet97},
evaluate the buying and selling decisions at a time step, regardless
of the need to update the historical effects of the previous decisions.
In other models, one-step expectations of the agents are considered,
leading to the \$-game \cite{Andersen03}. 
There are also market models with a mixture
of trend-following and fundamentalist agents \cite{Marsili01, Demartino03} 
or markets with crossover regimes 
dominated by trend-following and fundamentalist strategies 
\cite{Lux99, Demartino04}. 
As pointed out in \cite{Yeung08}, 
these models do not reflect the history-dependent
considerations of real market agents.

Improved versions of the minority games considered agents 
using {\it virtual wealth}
to evaluate their strategies \cite{Challet08}. The Wealth Game (WG)
\cite{Yeung08} was introduced to overcome this deficiency. Agents
in WG evaluate their strategies by calculating their virtual wealth,
that is, the wealth (including cash and stocks) that the strategies
would bring were their recommendations completely adopted in history.
The most significant advantage of this evaluation method can be seen
when agents are allowed to make {\it holding} decisions (that is,
decisions to take no buying or selling actions). In WG, the virtual
wealth due to holding long (short) positions increases when stock
prices are rising (dropping). On the other hand, virtual point updates
in the original MG are neutral to holding positions.
Tests with real market and artificial data 
confirm the versatility of wealth-based strategies \cite{Yeung08, Baek10}. 

A consequence of using wealth-based strategies in WG is the emergence
of price cycles through the self-organization of the different types
of agents. This is both an important and interesting issue, since
it sheds light on the formation and disappearance of bubbles and crashes
in real financial markets. Giardina and Bouchaud considered a model
with bubbles and crashes in the price trend of the market \cite{Giardina03}.
The behavior was explained in terms of the interplay between the 
trend-following 
and fundamentalist behaviors of the agents, but the mechanism of the
disappearance of this periodic phase remains an open issue. In WG,
the roles of the trendsetters and fickle agents in sustaining the
price cycles were explained, and the disappearance of the periodic
phase was attributed to the failure of the trendsetters to gain wealth
from the fickle agents. (The trendsetters are synonymous to the 
trend-followers in the literature, but are renamed trendsetters
to emphasize their role in initiating the bubbles and crashes.) However,
this picture assumes the presence of market makers, who manage to
fulfill buy and sell orders irrespective of the order imbalance. 
This is not applicable to the stock market, 
since in the absence of market makers, the market clearing mechanism requires
an exact matching of buy and sell orders. 
Consequently, not all agents can have 
their buying or selling orders fulfilled. 
Thus, the unfulfilled buyers
(sellers) would repeat their bids (asks) step after step. This creates
a much stronger tendency for the price to go monotonically upwards
(downwards). The appearance of the periodic phase becomes questionable
and, if it exists, its mechanism of formation and disappearance may
not necessarily be the same.

In this paper, we study a simplified model of WG in the absence of
market makers, focusing on the mechanism creating and destroying the
cycles of bubbles and crashes. We will adopt a minimalist approach,
and consider the simple but essential elements that contribute to
the studied mechanism. It turns out that with trend-followers and
fundamentalists of memory size 2 being the two main groups of investors,
the price dynamics already exhibits many interesting features. (The
fundamentalists are further divided into optimistic and pessimistic
subgroups.) On one hand, these groups are inclusive enough to represent
the attitudes of most investors, and on the other hand, simple enough
to enable convenient analyses. Despite the simplification, we will
see that many important features and phase transitions in the original
WG with market makers are preserved. Rich econophysical implications
are revealed regardless of the simplifications.

This paper is outlined as follows. After introducing the model in
Sec. \ref{sec:Introduction}, we describe different attractor 
behaviors in the space of price sensitivity and market impact in Sec.
\ref{sec:Phase-Diagram}. Analytical studies about the phase transitions,
including the cause of the transition and the precise location of
the phase boundary, are discussed in Sec. \ref{sec:The-Transient-Period}
to \ref{sec:The-Phase-Boundary}. In Sec. \ref{sec:f-dependence},
we study the dependence of the attractor behavior on the fraction
of trend-followers, accompanied by a concise analytical study about
the corresponding phase transition. Finally, the conclusion is drawn
in Sec. \ref{cha:Conclusion}.

\section{\label{sec:Introduction}The Model}

The Wealth Game \cite{Yeung08} consists
of $N$ agents playing in a single-commodity market. For convenience,
we will use the language of stock markets in the following discussions.
At each time step, the agents make decisions to buy, sell, or hold (no action)
stocks, based on the predictions of their best strategies. The decision
of agent $i$ at time $t$ is denoted as $a_{i}(t)=1,-1,0$, which
corresponds to buy, sell or hold respectively. A strategy takes the
the signs of previous $m$ historical price changes (represented by
a string of $\uparrow$ and $\downarrow$) as the input signal, and
the output signal is the advice on the trading action of the present
step. Table \ref{tab:typical strategy} shows the possible content
of a strategy for $m=2$. We require each usable strategy to have
at least one buying and one selling prediction, or else it is too
dull to be used. With this restriction, $s$ strategies are randomly
drawn to each agent. 

\begin{table}
\caption{\label{tab:typical strategy}The content of a typical strategy for
$m=2$.}
\noindent \begin{centering}
\begin{tabular}{|c|c|}
\hline 
Input signal & Output Signal (advice)\tabularnewline
\hline
\hline 
$\uparrow\uparrow$ & Buy\tabularnewline
\hline 
$\uparrow\downarrow$ & Sell\tabularnewline
\hline 
$\downarrow\uparrow$ & Hold\tabularnewline
\hline 
$\downarrow\downarrow$ & Sell\tabularnewline
\hline
\end{tabular}
\par\end{centering}

\end{table}

The position of agent $i$ at time $t$ is given by
\begin{equation}
	k_{i}(t)={\displaystyle \sum_{t'=0}^{t}}a_{i}(t'),
\label{eq:position}
\end{equation}
which records the number of stocks possessed by an agent. Short selling
is allowed, such that $k_{i}(t)$ can be negative. We assume that
each agent has limited assets, so the restriction $|k_{i}(t)|\leq K$
is applied, i.e. actions that further increase $|k_{i}(t)|$ to exceed
$K$ are ignored, and so $K$ denotes the maximum number of stocks
that an agent can buy or short sell. The market price evolves in response
to the market's excess demand $A(t)$, which is defined as\begin{equation}
A(t)=\sum_{i=1}^{N}a_{i}(t),\label{eq:excess demand}\end{equation}
and the price is updated by
\begin{equation}
	P(t+1)=P(t)+\mbox{sgn}[A(t)]|A(t)|^{\gamma},
\label{eq:price update}\end{equation}
where $\gamma\in[0,1]$ is the market sensitivity controlling how
sensitively the price changes with the market excess demand. $\gamma=0$
corresponds to a step function \cite{Challet97, Savit99, Jefferies01} 
while $\gamma=1$ corresponds to a
linear function \cite{Challet00, Marsili00, Challet00b, Heimel01}. 
Suppose an agent would like to buy a stock at price
$P(t)$. Then she queues up in the market to wait for her turn of
a transaction. Depending on how long the queue is, the actual transaction
price $P_{\mathrm{T}}(t)$ may deviate from her desired price $P(t)$.
This is one of the examples showing how the market impact (i.e. the
collection of all the market factors imposed by agents' participation)
would influence the agents' trading activities \cite{Challet08}.
In this model, the transaction price is defined as
\begin{equation}
	P_{\mathrm{T}}(t)=(1-\beta)P(t)+\beta P(t+1),
\label{eq:transaction price}\end{equation}
where $\beta\in[0,1]$ is the market impact. For convenience, we assume
that all the agents are affected to the same extent by market impact.
If $\beta=0$, the market impact is small so that the agent can immediately
trade with her most desired price. When $\beta=1$, the queue is so
long (and the market impact is so large) that the agent is actually
trading with $P(t+1)$, which may have already deviated considerably
from $P(t)$.

The wealth of an agent consists of two parts: cash in her hand and
the value of stocks she is holding. Agents' cash is updated by
\begin{equation}
	c_{i}(t)=c_{i}(t-1)-a_{i}(t)P_{\mathrm{T}}(t),
\label{eq:cash update}\end{equation}
while the agents' wealth at the moment they just finish the transactions
at time $t$, is defined as
\begin{equation}
	w_{i}(t)=c_{i}(t)+k_{i}(t)P_{\mathrm{T}}(t).
\label{eq:agent wealth}\end{equation}

Among the $s$ strategies, an agent only chooses one to follow at
each time step. The virtual position, cash and wealth of a strategy
will be calculated in the same way as an agent, by Eqs. (\ref{eq:position})
(a strategy is also restricted by $K$), (\ref{eq:cash update}) and
(\ref{eq:agent wealth}). Its virtual wealth evolves when its prediction
is applied in the market. The best strategy is then defined as the
one with the highest accumulated virtual wealth, which is to be adopted
by the agent. When a previously best strategy is outperformed, switching
strategies by agents occurs.

\subsection{Market without Market Makers}

The original Wealth Game implicitly assumes the participation of market
makers. This means that when there are more buyers than sellers, market
makers will provide stocks to the excess buyers, and when there are
more sellers than buyers, they will absorb the extra stocks. Withdrawing
the market makers from the game implies that the supply and demand
cannot be balanced. To achieve a balance, an apparent way is to randomly
pick some excess majority traders and ignore their orders. For the
sake of fairness, however, we assume that all the majority traders
reduce their orders such that all of them can only be partially satisfied
\cite{Giardina03, Caldarelli97, Slanina99}. 

The mathematical modification to the original game is as follows.
The quotation of agent $i$ ($j$) who wants to buy (sell) is defined
as
\begin{equation}
	q_{i}^{\mathrm{buy}}=\mathrm{min}(1,K-k_{i}),
\label{eq:buying quotation}\end{equation}
\begin{equation}
	q_{j}^{\mathrm{sell}}=-\mathrm{min}(1,K+k_{j}).
\label{eq:selling quotation}\end{equation}
The quotation is the amount of stock an agent wants to trade. It is
defined this way since the agents can now buy (sell) a fraction of
their original units of stock, and the stocks held (short sold) by
each agent are still required to be bounded by the maximum position
$K$. We define the sum of buying (selling) quotations as
\begin{equation}
	A_{\mathrm{buy}}=\sum_{i}q_{i}^{\mathrm{buy}},
\label{eq:buying sum}\end{equation}
\begin{equation}
	A_{\mathrm{sell}}=\sum_{j}q_{j}^{\mathrm{sell}}.
\label{eq:selling sum}\end{equation}
The modification to the excess demand $A(t)$ for Eq. (\ref{eq:excess demand})
is
\begin{equation}
	A(t)=\sum_{l=1}^{N}q_{l}(t)=A_{\mathrm{buy}}+A_{\mathrm{sell}}.
\label{eq:mod excess demand}\end{equation}
When $A(t)$ is positive, the position change of a buying (selling)
agent after each transaction is
\begin{equation}
	\Delta k_{i}^{\mathrm{buy}}=\left(\frac{|A_{\mathrm{sell}}|}
	{A_{\mathrm{buy}}}\right)q_{i},\label{eq:del k_buy}
\end{equation}
\begin{equation}
	\Delta k_{j}^{\mathrm{sell}}=q_{j}.
\label{eq:del k_sell}\end{equation}
When $A(t)$ is negative,
\begin{equation}
	\Delta k_{i}^{\mathrm{buy}}=q_{i},
\label{eq:del k_buy2}\end{equation}
\begin{equation}
	\Delta k_{j}^{\mathrm{sell}}=\left(\frac{A_{\mathrm{buy}}}
	{|A_{\mathrm{sell}}|}\right)q_{j}.\label{eq:del k_sell2}
\end{equation}
Based on the above modifications, one could easily verify that the
supply and demand can be balanced at any time step. Note that, now
the market price change is solely driven by agents' bid-ask actions,
regardless of whether transactions are really carried out afterwards.
This is the cause of so called unfulfilled orders \cite{Giardina03},
``dry quoting''. In this case, there is only one quoting group (e.g.
the bidding group) who cannot find their matching dealers. This essentially
resembles the circumstance of a market when the price fluctuation
is significant but the trading volume is negligible. It should also
be emphasized that the absence of market makers implies that the market
is zero-sum, which means the total wealth of all the agents is conserved,
as the gain of an agent must be accompanied by the loss of another
agent.

For the updating of the virtual positions of the strategies, one may
either use the original scheme of $\Delta k_{\xi}=\pm1$ subject to
the constraints of the maximum and minimum positions $\pm K$, or
use the modified scheme analogous to Eqs. (\ref{eq:buying quotation}),
(\ref{eq:selling quotation}), (\ref{eq:del k_buy}) - (\ref{eq:del k_sell2}).
In this paper we use the original scheme, and have checked that both
schemes yield qualitatively similar results.

\subsection{Cash Rule}

It was found that merely withdrawing the market makers from the game
only creates uninteresting market dynamics. Due to the imbalance between
supply and demand, majority traders can only be partially satisfied.
Time step after time step, they quote again and again, boosting (busting)
the price monotonically and creates an ever increasing (decreasing)
trend. 

To get rid of this undesirable feature in our model, we need to take
into account the inability of the agents to order when the stock price
is too high or low. Hence we propose that the agents are forced to
cease their quotations if the following conditions are satisfied:\begin{equation}
c_{i}^{\mathrm{buy}}(t)-P(t)<0,\label{eq:cash rule1}\end{equation}
\begin{equation}
c_{j}^{\mathrm{sell}}(t)+P(t)<0,\label{eq:cash rule2}\end{equation}
where the superscripts \textsl{buy} and \textsl{sell} stand for
buyers and sellers respectively. Condition (\ref{eq:cash rule1})
is essentially saying that agents who have too little cash would not
bother to go for queuing if the price is too high. Condition (\ref{eq:cash rule2})
means that if the price is too low, agents who are not liquid enough
would not take the risk to borrow stocks.

\subsection{Strategies: Fast Trendsetters, Top and Bottom Triggers}

To facilitate analyses, the strategy space of the game is simplified
in the following ways. First, the outputs of a strategy are restricted
to buying and selling only. No holding actions are included. An agent
stops quoting only when she tries to place a buying (selling) quotation
but the restriction $|k_{i}|=K$ is reached, or when she is restricted
by the Cash Rule (i.e. conditions (\ref{eq:cash rule1}) 
or (\ref{eq:cash rule2})
are satisfied). The direct consequence is that the total number of
possible strategies becomes $2^{2^{m}}$ instead of $3^{2^{m}}$.
Second, we only study the special case where an agent has two strategies
($s=2$) and considers two historical price changes as the strategy
input signal ($m=2$). Hence there are $2^{2^{m}}=16$ strategies.
We further focus on those strategies with opposite decisions for inputs
$\uparrow\downarrow$ and $\downarrow\uparrow$. This reduces the
set of strategies to those listed in Table \ref{tab:F, T, B strategy content}
and their antistrategies. They are called fast trendsetters (F), top
trigger (T), bottom trigger (B) and slow trendsetters (S) respectively.
The meanings of these names will become clear in the next paragraphs.
Furthermore, studies in the original Wealth Game shows that S strategy
plays similar role as the F strategy in the formation of price cycles.
Hence, we restrict the strategy space, and only three strategies are
\textsl{evenly} assigned to the agents, namely, the F, T and B strategies.

\begin{figure}
\noindent \begin{centering}
\includegraphics[width=86mm]{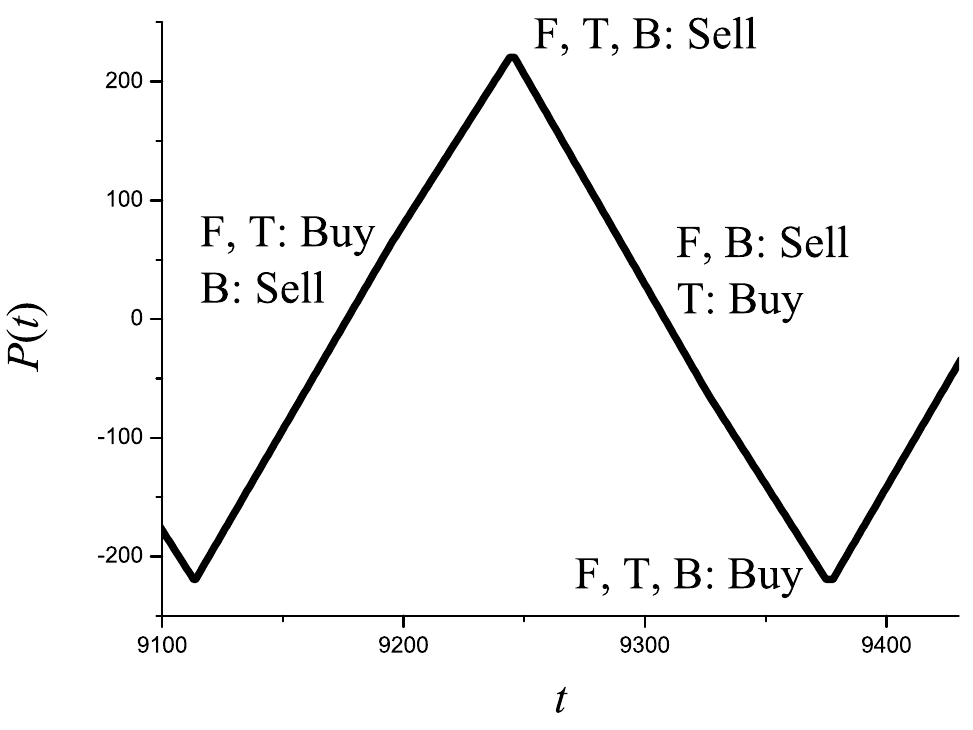}
\par\end{centering}

\caption{\label{fig:artificial trendy market}
The price series of an artificial
trendy market. Different strategies make different decisions.}

\end{figure}

We now consider the outputs of these three strategies in an artificial
trendy market, as shown in Fig. \ref{fig:artificial trendy market}. 

\begin{table*}
\caption{\label{tab:F, T, B strategy content}The F, T, B and S strategies.}
\noindent \begin{centering}
\begin{tabular}{|c|c|c|c|}
\hline 
F strategy & T strategy & B strategy & S strategy\tabularnewline
\hline
\hline 
\begin{tabular}{c|c}
Input & Output\tabularnewline
\hline
\hline 
$\uparrow\uparrow$ & Buy\tabularnewline
\hline 
$\uparrow\downarrow$ & Sell\tabularnewline
\hline 
$\downarrow\uparrow$ & Buy\tabularnewline
\hline 
$\downarrow\downarrow$ & Sell\tabularnewline
\end{tabular} & \begin{tabular}{c|c}
Input & Output\tabularnewline
\hline
\hline 
$\uparrow\uparrow$ & Buy\tabularnewline
\hline 
$\uparrow\downarrow$ & Sell\tabularnewline
\hline 
$\downarrow\uparrow$ & Buy\tabularnewline
\hline 
$\downarrow\downarrow$ & Buy\tabularnewline
\end{tabular} & \begin{tabular}{c|c}
Input & Output\tabularnewline
\hline
\hline 
$\uparrow\uparrow$ & Sell\tabularnewline
\hline 
$\uparrow\downarrow$ & Sell\tabularnewline
\hline 
$\downarrow\uparrow$ & Buy\tabularnewline
\hline 
$\downarrow\downarrow$ & Sell\tabularnewline
\end{tabular} & \begin{tabular}{c|c}
Input & Output\tabularnewline
\hline
\hline 
$\uparrow\uparrow$ & Buy\tabularnewline
\hline 
$\uparrow\downarrow$ & Buy\tabularnewline
\hline 
$\downarrow\uparrow$ & Sell\tabularnewline
\hline 
$\downarrow\downarrow$ & Sell\tabularnewline
\end{tabular}\tabularnewline
\hline
\end{tabular}
\par\end{centering}

\end{table*}

A glance at the content of the F strategy suggests that it is a trend-believing
strategy. The F strategy advises to buy in a rising trend, sell at
the price peak, sell in a falling trend and buy at the price valley.
Once a sufficiently long price trend (either rising or falling) has
been established, the F strategy gains most wealth and hence would
be adopted by most agents. If this trend-believing strategy is adopted
by the majority, the price trends can be set up persistently. In the
literature, it is called the trend-follower 
\cite{Lux99, Farmer99, Jefferies01, Marsili01, Andersen03, 
Giardina03, Demartino03, Demartino04}.
Here, to highlight its role in perpetuating the price cycles, it is
called the ``trendsetter''. It is {}``fast'' as agents adopting
it react immediately to reversals in the price trend in a trendy market,
in contrast to the slow trendsetters who join the trendsetting bandwagon
one step slower, according to the outputs prescribed 
in Table \ref{tab:F, T, B strategy content}
for inputs $\uparrow\downarrow$ and $\downarrow\uparrow$.

The T and B strategies have a fundamentalist character. The T strategy
gives buying advice in response to all inputs, except when the price
trend reaches a peak, that is, when the signal is $\uparrow\downarrow$.
Combined with the positions bounds, an agent following its advice
prefers to stay in a long position. Therefore it can be considered
as an optimistic fundamentalist. Fundamentalists believe that the
price should not deviate from a fundamental value. When the price
is lower, they try to stick to long positions. The T strategy is optimistic
because it targets a relatively high fundamental value signaled by
a price peak. In the original Wealth Game, the selling actions advised
by the T strategies help to trigger the falling trends in price cycles.
Hence they are called the {}``top trigger''.

Similarly, the B strategy gives selling advice in response to all
inputs, except when the input is $\downarrow\uparrow$. It is pessimistic
since it prefers to stick to a short position and buys only when the
price reaches a valley.

In summary, we are studying a model in which each agent is equipped
with two strategies; each of them may either be a trend-following
(F) strategy or a fundamentalist one (either optimistic (T) or pessimistic
(B)). Note that the virtual wealth of a strategy is not influenced
by the Cash Rule and the absence of market makers, meaning that an
agent evaluates it solely according to the original Wealth Game (Eqs.
(\ref{eq:position}), (\ref{eq:cash update}) and (\ref{eq:agent wealth}))
and assumes that the virtual transaction of a strategy is always successful.
We also note that the game is invariant under the gauge transformation
mapping the $\uparrow$ and $\downarrow$ signals to each other and
the buy and sell decisions to each other.

\section{\label{sec:Phase-Diagram}Phase Diagram}

We study the behavior of the game by starting from a set of unbiased
initial conditions. Each agent has the same initial cash $c_{\mathrm{o}}$
and do not hold any stocks initially ($k_{i}\left(0\right)=0$), so
that the initial wealth of each agent is $w_{i}\left(0\right)=c_{o}$.
The initial stock price is $P\left(0\right)=0$, and the initial virtual
wealth of all strategies $\sigma$ is $w_{\sigma}\left(0\right)=0$.
To avoid ambiguous decisions when more than one strategies have the
same virtual wealth during the game, one of the $3!$ priority orderings
of F, T and B is randomly chosen before the game starts. We call this
{}``throwing the public dice''. To initiate the game dynamics, one
of the four historical price signals ($\uparrow\uparrow$, $\uparrow\downarrow$,
$\downarrow\uparrow$ or $\downarrow\downarrow$) is randomly chosen.
Considering the choices in the public dice and the price signal, there
are 24 possible sets of initial conditions. Due to the gauge symmetry
in the game, there are 12 distinct initial conditions. 

\begin{figure}
\noindent \begin{centering}
\includegraphics[width=86mm]{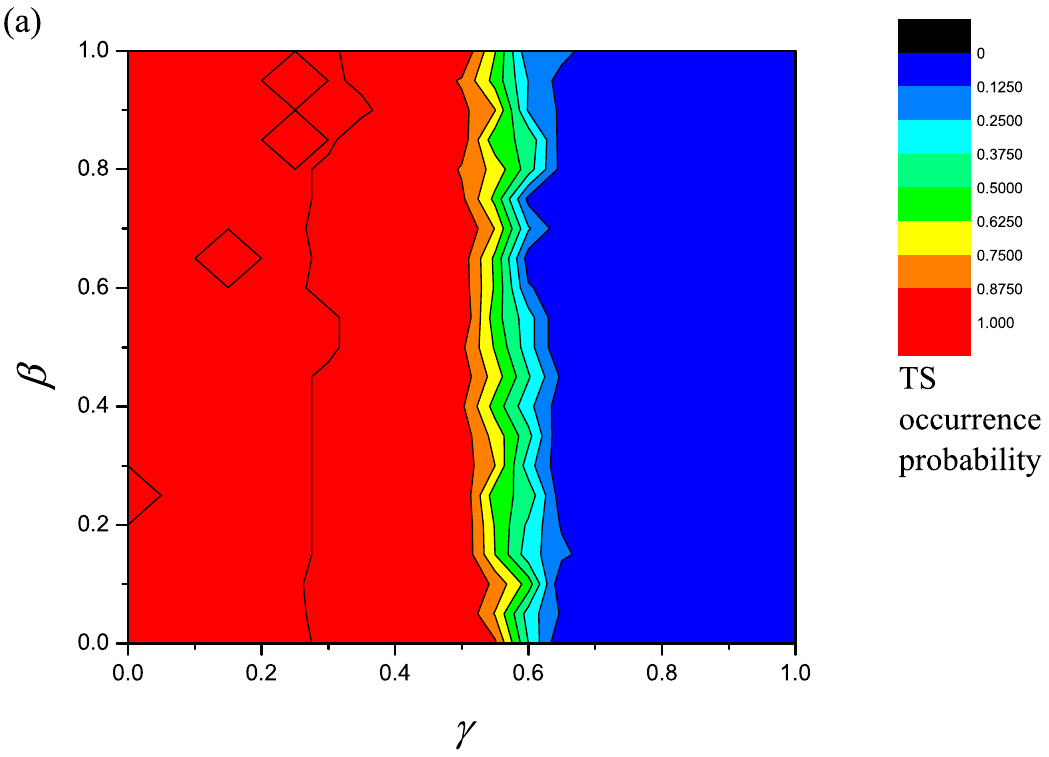}
\par\end{centering}

\noindent \begin{centering}
\includegraphics[width=86mm]{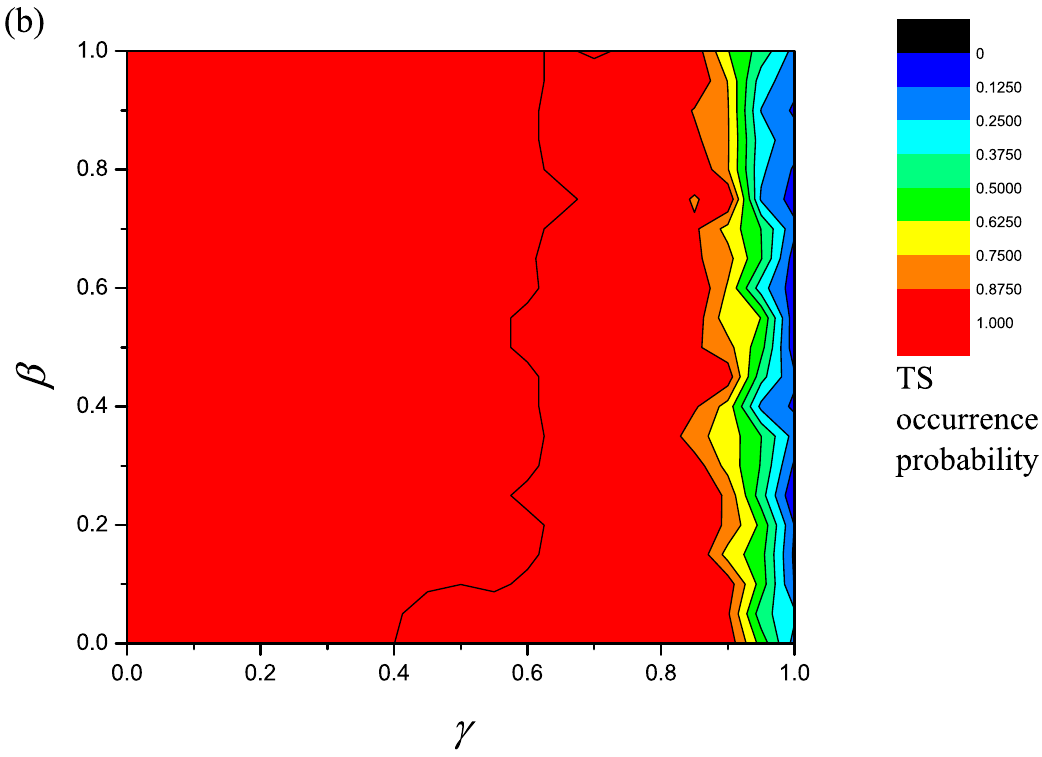}
\par\end{centering}

\caption{\label{fig:phase diagram1}(Color online) The phase diagram in the space of price
sensitivity $\gamma$ and market impact $\beta$ for (a) $c_{\mathrm{o}}=100$,
and (b) $c_{\mathrm{o}}=925$. Other parameters are $m=2$, $s=2$,
$K=3$, $N=1000$ (20 samples).}

\end{figure}

\begin{figure}
\noindent \begin{centering}
\includegraphics[width=86mm]{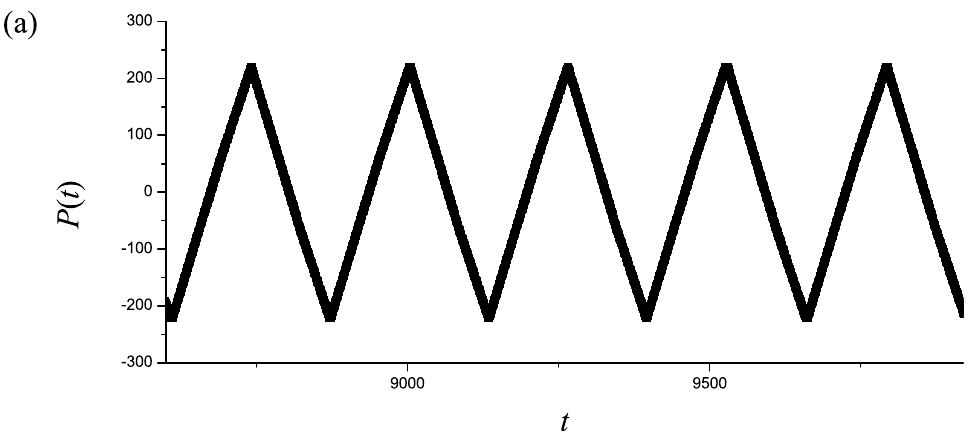}
\par\end{centering}

\noindent \begin{centering}
\includegraphics[width=86mm]{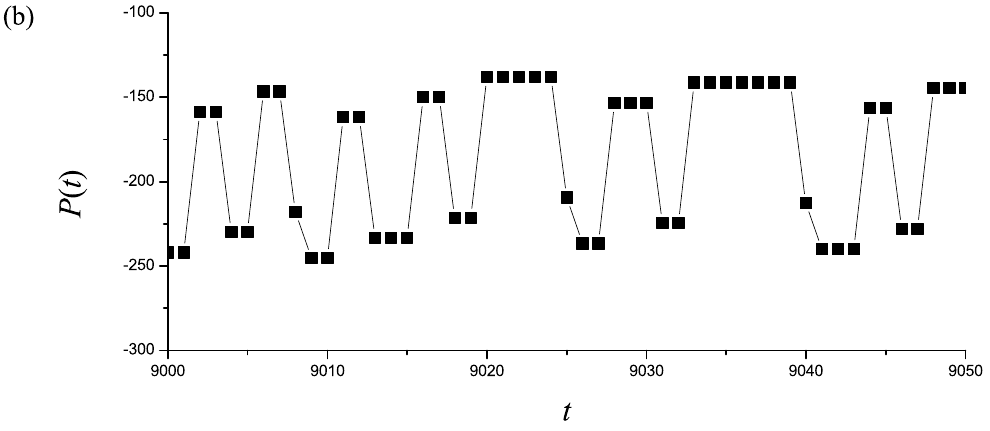}
\par\end{centering}

\caption{\label{fig:TS NTS time series}The time dependence of price for a
typical sample with $m=2$, $s=2$, $K=3$, $N=1000$, $c_{\mathrm{o}}=100$
for (a) the trendsetters' phase at $(\gamma,\beta)=(0.2,0.2)$, (b)
the bouncing phase at $(\gamma,\beta)=(0.7,0.2)$.}

\end{figure}

\begin{figure}
\noindent \begin{centering}
\includegraphics[width=86mm]{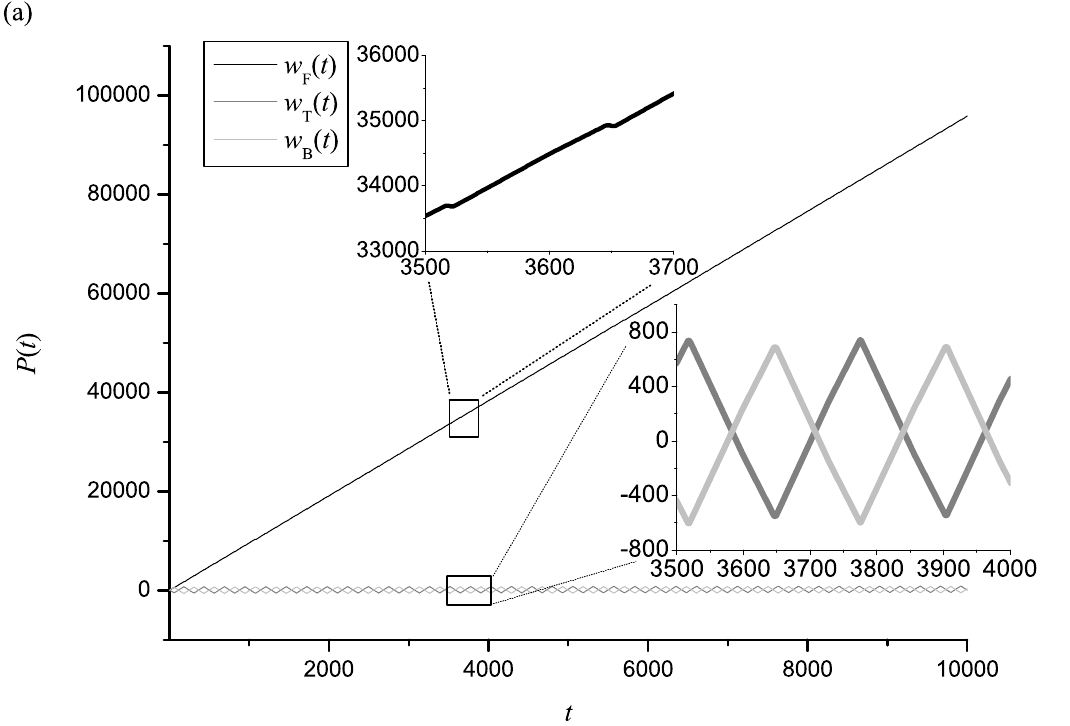}
\par\end{centering}

\noindent \begin{centering}
\includegraphics[width=86mm]{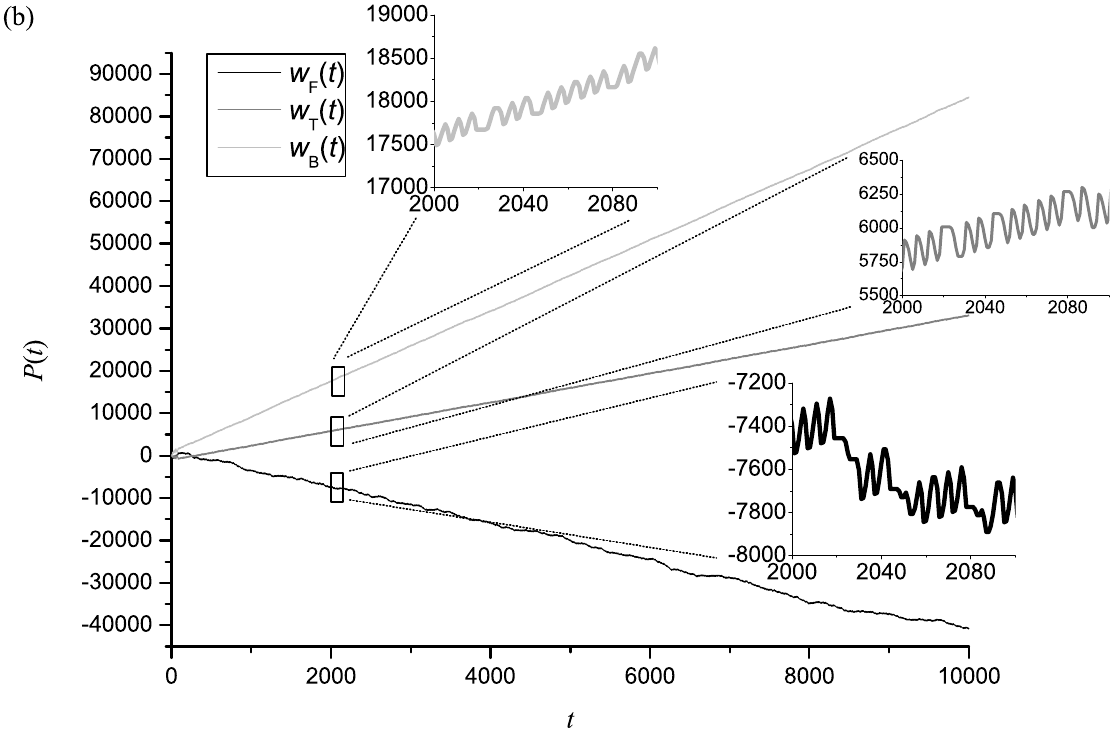}
\par\end{centering}

\caption{The virtual wealth of the F, T and B strategies in (a) the TS attractor
at $(\gamma,\beta)=(0.2,0.2)$, (b) the BO attractor at $(\gamma,\beta)=(0.7,0.2)$.
Parameters: $m=2$, $s=2$, $K=3$, $N=1000$, $c_{\mathrm{o}}=100$.\label{fig:TS BO virtual wealth}}

\end{figure}

The steady state behavior of the game can be summarized in the phase
diagram in Fig. \ref{fig:phase diagram1} in the space of price
sensitivity $\gamma$ and market impact $\beta$. The space is divided
into two phases: the trendsetters' (TS) phase at low $\gamma$ and
the bouncing (BO) phase at high $\gamma$. 
When the initial cash $c_{\mathrm{o}}$
increases, the TS phase expands. We also observe that the phase transition
is weakly dependent on the market impact $\beta$, 
which implies that the market is dominated by dry quoting. 
Typical time series
of the price in the attractors of these phases are shown in Fig.
\ref{fig:TS NTS time series}. The corresponding virtual wealth of
the F, T and B strategies are shown in Fig. \ref{fig:TS BO virtual wealth}.

We remark that the phase transition in Fig. \ref{fig:phase diagram1}
is a dynamical rather than a generic transition. The occurrence and
the position of the transition is specific to the unbiased initial
condition described above. Starting with other initial conditions,
or introducing perturbations to the dynamics, 
we can obtain
the bouncing attractor in the trendsetters' phase and vice versa.
For example, we have done numerical experiments by starting from the
TS phase and gradually increasing $\gamma$ until we reach the BO
phase, but we observe that the TS attractor remains stable. Similar
annealing experiments from the BO phase to the TS phase also show
that the BO attractor can be stable in the TS phase. 

In another set of experiments, we first prepare the TS attractor in
the TS phase, and then inject virtual cash to the B strategy so that
the BO attractor is favored. We observe that for all values of $\gamma$,
stable BO attractors can be formed when the amount of injected virtual
cash is sufficiently large. In the converse experiment, we inject
virtual cash to the F strategy in the BO attractor in the BO phase.
When $c_{\mathrm{o}}$ is high so that the BO phase is narrow, stable
TS attractors are formed at all values of $\gamma$ when the level
of injected virtual cash is sufficiently high. However, when $c_{\mathrm{o}}$
is low and the BO phase is broad, we found that at high values of
$\gamma$, stable TS attractors cannot be formed no matter how much
virtual cash is injected. This indicates that the conditions for forming
the TS attractor are probably more restrictive than the BO attractor,
and the existence phase of the TS attractor may be studied using approaches
to generic transitions. However, this transition is not relevant to
the dynamical transitions starting from the unbiased initial conditions
discussed in this paper. 

\subsection{The Bouncing Phase}

In the bouncing phase, the dominant strategy is either T or B. In
the example shown in Fig. \ref{fig:TS NTS time series}(b), the dominant
strategy is B. Its dynamics is one large upward jump in price, followed
by one or several steps of zero change, and a downward jump in several
small steps or one large step. It reflects the desire of the agents
who buy stocks to take advantage of a low price, but the price trend
is prevented from rising further due to the prevailing pessimistic
atmosphere of the market.

The virtual wealth of the strategies are shown 
in Fig. \ref{fig:TS BO virtual wealth}(b).
Since the B strategy gains profit from buying immediately before the
large upward jump and selling afterwards, it becomes the dominant
strategy. The T strategy also gains profit from selling immediately
before the downward jumps and buying at a lower price. Hence its virtual
wealth also increases with time, but at a rate slower than that of
the B strategy. On the other hand, since the F strategy cannot gain
wealth in a price series with frequent trend reversals, it becomes
a losing strategy.

Besides the up-bouncing example shown in Fig. \ref{fig:TS NTS time series}(b),
down-bouncing dynamics dominated by the T strategy can also be observed,
depending on the initial conditions. Using the gauge symmetry of the
game, their mechanism can be explained similarly. 

\subsection{The Trendsetters' Phase}

This is the phase that gives rise to bubbles and crashes in the cycles
of stock prices \cite{Giardina03, Yeung08}. 
As shown in Fig. \ref{fig:TS NTS time series}(a),
the price series consists of alternating long rising and falling trends.
As shown in Fig. \ref{fig:TS BO virtual wealth}(a), the virtual wealth
of the F strategy rises continuously due to its trendy decisions,
with minor setbacks when the price trend reverses. On the other hand,
the T and B strategies oscillate about their average at the
frequency of the price oscillations. Since the T strategy mainly takes
a long position, its virtual wealth is higher than that of the B strategy
during the upper half of the price cycle. Similarly, the B strategy
has a higher virtual wealth during the lower half of the price cycle.
The T strategy gains wealth by selling at the peak and buying back
at a lower price the next step, and the B strategy gains wealth by
buying at the price minimum and selling at a higher price at the next
step. Hence their average also has a slowly rising trend.

In the absence of market makers, the steady state is dominated by
dry quotations. Hence the cash of different agents becomes stationary
when the system equilibrates. For $s=2$, there are six types of agents.
Those holding two F strategies are denoted as FF agents, and the other
five types are FT, FB, TT, TB and BB agents. In the steady state,
agents can be roughly categorized as the active groups (including
FF, FT and FB agents) and the inactive group (including TT, TB and
BB agents). The active group usually has a relatively large amount
of cash, so that the price movement is mostly due to their participation
in the quoting activities. The inactive agents usually have little
cash, or have already reached the position bounds $(|k_{i}|=K)$,
so that their quotations would usually be terminated by the Cash Rule
or the position bounds. Due to the unavailability of participation
of the inactive group, real transactions cannot materialize as the
active agents cannot find matching dealers. Thus in this phase, the
price movement is solely caused by dry quoting. 

\begin{figure}
\noindent \begin{centering}
\includegraphics[width=86mm]{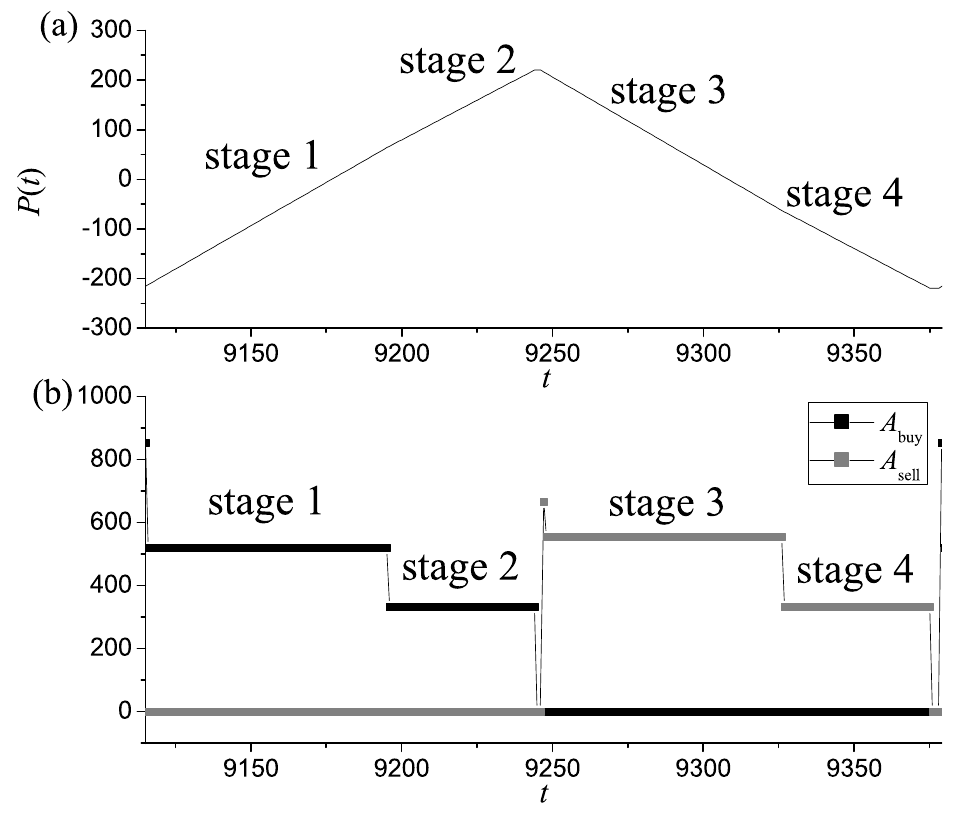}
\par\end{centering}

\caption{\label{fig:TS series}A typical trendsetters' attractor showing (a)
the price evolvement, (b) the evolvements of the total buying and
selling quotations. The parameters are $(\gamma,\beta)=(0.2,0.2)$,
$m=2$, $s=2$, $K=3$, $N=1000$, $c_{\mathrm{o}}=100$.}

\end{figure}

The TS attractor is generally divided into four stages as shown in Fig.
\ref{fig:TS series}. In stage 1, all active agents place buying quotes
and thus the price is boosted up with the steepest slope. Stage 2
starts when the price has reached a level higher than the cash level
of some of the active agents, whose quoting actions are terminated
by the Cash Rule, leading to a decrease in the rate of price evolvement.
Stage 2 ends when the price is higher than the cash level of the most
liquid active agents. A quiet step (i.e. a time step with zero price
change) follows as no one quotes. Since the price change is zero,
a random signal is generated as the signal input for the next step.
If $\uparrow$ is generated, another quiet step follows, until a $\downarrow$
signal is randomly generated, triggering a falling trend. Stages 3
and 4 are duplicates of stages 1 and 2 under the gauge transformation.

In Fig. \ref{fig:TS series}, note that the transition from stage
1 to 2 (or from stage 3 to 4) is not apparent in the price series.
This is because TS attractors exist at low $\gamma$, where the price
change is weakly sensitive to the excess demand $A(t)$.

In the following sections, we will analyze the behavior of the TS attractor,
leading eventually to an estimation of the phase transition point.

\section{The Transient Period\label{sec:The-Transient-Period}}

In contrast to the trendsetter attractor in \cite{Yeung08}, the TS
attractor in the absence of market makers is dominated by dry quotations.
This implies that the amount of cash held by each agent becomes stationary
at the steady state. As will be confirmed in the next section,
the period of the price cycles depends on the cash level of the most
liquid agent. Hence, the steady state behavior of the price cycles
directly depends no how the transient dynamics redistributes the cash
into the hands of the different types of agents. This heavy dependence
on the transient dynamics is the characteristics of the Wealth Game
without market makers. 

\begin{figure}
\noindent \begin{centering}
\includegraphics[width=86mm]{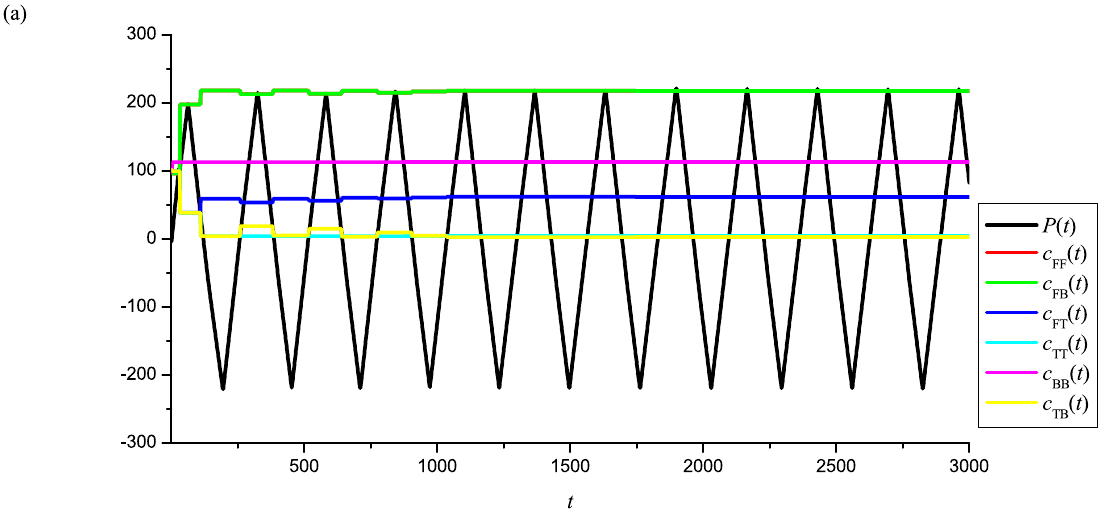}
\par\end{centering}

\noindent \begin{centering}
\includegraphics[width=86mm]{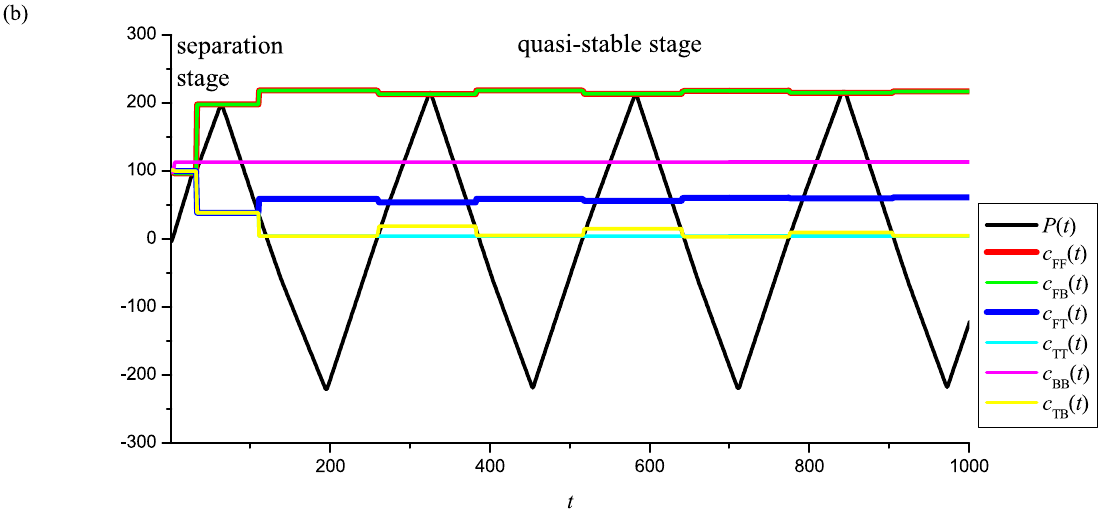}
\par\end{centering}

\noindent \begin{centering}
\includegraphics[width=86mm]{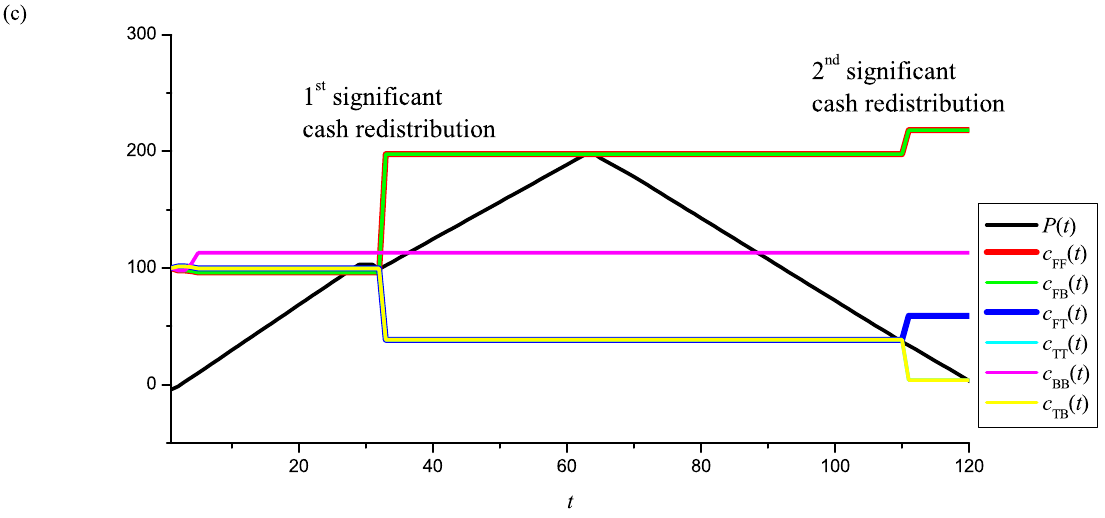}
\par\end{centering}

\caption{(Color online) The cash evolvements of different agents and the price series in type
I TS attractor for (a) the asymptotic time series, (b) the transient,
which is further divided into the quasi-stable stage and the separation
stage, and (c) the separation stage. 
The parameters are $(\gamma,\beta)=(0.2,0.8)$,
$m=2$, $s=2$, $K=3$, $N=1000$, $c_{\mathrm{o}}=100$, initial
conditions ($\downarrow\downarrow$, $\mathrm{B}>\mathrm{F}>\mathrm{T}$).
\label{fig:rank 1 transient}}

\end{figure}

A typical price series is shown in Fig. \ref{fig:rank 1 transient}
when the initial cash $c_{\mathrm{o}}$ is sufficiently large that
we are not very close to the phase boundary. At the beginning of the
transient stage, a significant redistribution of cash starts to take
place in the first half of the quasi-period of the price series. We
refer to this as the {\it separation stage}. Following the separation
stage, the cash levels are roughly flat with occasional jumps, whose
magnitudes are progressively smaller. 
We call this the {\it quasi-stable stage}.
Eventually, the cash levels become stable at the steady state.

At the first few steps of the separation stage, the price rises from
zero. Some real transactions take place, but since the price is close
to zero, the cash levels of the agents do not change much and are
roughly equal to $c_{\mathrm{o}}$. After several steps, the major
sellers (the BB agents) have reached the minimum position of $-K$.
Dry quoting happens again and again, until the price is just higher
than $c_{\mathrm{o}}$, at where the agents think the stock is too
expensive. The price stops rising and when a downward signal appears,
all agents make selling quotations. In the next step, the signal $\downarrow\downarrow$
first appears after the quiet period. Both buying and selling quotations
are made by the agents. This is the time when there is a significant
redistribution of cash among the agents.

Subsequent to this event, there is a rising or falling price trend
depending on the excess demand at the event. For attractors obtained
from the initial condition used in Fig. \ref{fig:rank 1 transient},
which will be classified as type I attractor, the price follows a
rising trend, hits a peak when it goes above the cash level of the
most liquid agents. No transactions can be fulfilled as the major
sellers (BB) have reached the minimum position already, and the price
goes on a falling trend. For other initial conditions, the price follows
a falling trend directly after the event. In all cases, such price
movements are not accompanied by any real transactions, as the price
is higher than the cash levels of the buying group. Hence, no cash
redistribution takes place for many time steps.

For type I attractors, the second significant cash redistribution
takes place when the price falls below the cash level of the second
most liquid group of agents. 

In the quasi-stable stage, real transactions can only materialize
when the price is close to zero, allowing the inactive agents to participate
with their low cash levels. When this happens, the cash levels of
the agents fluctuate little bit. These fluctuations are so minor that
we can neglect their effects on the final stable cash levels.

Hence we can focus on the cash evolvements in the separation stage 
to calculate the approximate stable cash levels and hence the amplitude
of the price cycles. For the initial condition 
in Fig. \ref{fig:rank 1 transient},
this is done with reference to Tables \ref{tab:first cash redis}
and \ref{tab:second cash redis} at the first and second significant
cash redistributions respectively. At the first event, the order of
priority of the strategies is T $>$ F $>$ B, leading to the adoption
of the strategies in the third column of Table \ref{tab:first cash redis}.
In response to the signal $\downarrow\downarrow$, the outputs of
these strategies are given in the fourth column. Taking into account
the positions listed in the fifth column, the final decisions of the
agents are given in the sixth column. Summing up the buying and selling
quotations weighted by the fractions in the second column, we obtain
$A_{\mathrm{buy}}=5N/9$ and $A_{\mathrm{sell}}=3N/9$. Thus, the
FF and FB agents are the minority and sell one whole unit of stock
at the price $P\approx c_{\mathrm{o}}$. Agents in the buying group
have bought $3/5$ unit of stocks, 
and have lost cash $\approx3c_{\mathrm{o}}/5$.

\begin{table*}
\caption{The outputs, positions, decisions and final cash (roughly in multiples
of $c_{\mathrm{o}}$) of the agents at the first significant cash
redistribution event of a type I TS attractor. \label{tab:first cash redis}}
\noindent \begin{centering}
\begin{tabular}{|c|c|c|c|c|c|c|}
\hline 
Agents & Fraction & Strategy & Output & Position & Decision & Final cash
\tabularnewline
\hline
\hline 
FF & $1/9$ & F & sell & $\geq-K+1$ & sell & 2\tabularnewline
\hline 
FT & $2/9$ & T & buy & $\leq K-1$ & buy & $2/5$\tabularnewline
\hline 
FB & $2/9$ & F & sell & $\geq-K+1$ & sell & 2\tabularnewline
\hline 
TT & $1/9$ & T & buy & $\leq K-1$ & buy & $2/5$\tabularnewline
\hline 
TB & $2/9$ & T & buy & $\leq K-1$ & buy & $2/5$\tabularnewline
\hline 
BB & $1/9$ & B & sell & $=-K$ & hold & 1\tabularnewline
\hline
\end{tabular}
\par\end{centering}

\end{table*}

\begin{table*}
\caption{The outputs, positions, decisions and final cash (roughly in multiples
of $c_{\mathrm{o}}$) of the agents at the second significant cash
redistribution event of a type I TS attractor. \label{tab:second cash redis}}
\noindent \begin{centering}
\begin{tabular}{|c|c|c|c|c|c|c|}
\hline 
Agents & Fraction & Strategy & Output & Position & Decision & Final cash\tabularnewline
\hline
\hline 
FF & $1/9$ & F & sell & $\geq-K+1$ & sell & $56/25$\tabularnewline
\hline 
FT & $2/9$ & F & sell & $\geq-K+1$ & sell & $16/25$\tabularnewline
\hline 
FB & $2/9$ & F & sell & $\geq-K+1$ & sell & $56/25$\tabularnewline
\hline 
TT & $1/9$ & T & buy & $\leq K-1$ & buy & 0\tabularnewline
\hline 
TB & $2/9$ & T & buy & $\leq K-1$ & buy & 0\tabularnewline
\hline 
BB & $1/9$ & B & sell & $=-K$ & hold & 1\tabularnewline
\hline
\end{tabular}
\par\end{centering}

\end{table*}

At the second event, the price has fallen to just below the cash level
of the second most liquid group of agents (FF, TT and TB), which is
roughly $2c_{\mathrm{o}}/5$. At this point, the price has fallen
to a value such that the order of priority of the strategies becomes
F $>$ T $>$ B. Consequently, as shown in Table \ref{tab:second cash redis},
the FT agents have changed to be in the selling group. We obtain $A_{\mathrm{buy}}=3N/9$
and $A_{\mathrm{sell}}=5N/9$. Hence, agents in the selling group
sell $3/5$ unit of stocks at the price $P\approx2c_{\mathrm{o}}/5$,
and so gain cash of the amount $6c_{\mathrm{o}}/25$. The buying agents
have their cash reduced by $2c_{\mathrm{o}}/5$.

As mentioned earlier, the maximum price $P_{\mathrm{max}}$ is reached
when the price has just exceeded the cash level of the most liquid
agents. Therefore, $P_{\mathrm{max}}\approx56c_{\mathrm{o}}/25$ for
type I attractors.

Considering the dynamics starting from all the 24 initial conditions,
we classify the TS attractors into three types as shown in Table \ref{tab:3 types attractors}.
Their maximum price and the distribution of the final cash are described
in Appendix. Although the amplitudes of the price cycles of the three
types of attractors are different, their dynamics are qualitatively
the same.

\begin{table}
\caption{The three types of TS attractors at $(\gamma,\beta)=(0.2,0.8)$, $m=2$,
$s=2$, $K=3$, $N=1000$, $c_{\mathrm{o}}=100$. In the initial conditions,
$*$ represents $\uparrow$ or $\downarrow$. Only the 12 initial
conditions that give rise to an initial rising trend during the separation
stage are included in the third column. The other 12 initial conditions,
related to those in the fourth column by gauge transformation $\uparrow\leftrightarrow\downarrow$
and T$\leftrightarrow$B, belong to the same corresponding type of
attractor. \label{tab:3 types attractors}}
\noindent \begin{centering}
\begin{tabular}{|c|c|c|c|}
\hline 
Attractor & Most liquid agents & Initial condition & $P_{\mathrm{max}}$\tabularnewline
\hline
\hline 
Type I & FF and FB & ($*$$*$, T $>$ F $>$ B) & $56c_{\mathrm{o}}/25$\tabularnewline
\hline 
Type II & FT & ($*$$\downarrow$, T $>$ B $>$ F) & 2$c_{\mathrm{o}}$\tabularnewline
\hline 
Type III & FB & ($*$$\uparrow$, T $>$ B $>$ F) & $8c_{\mathrm{o}}/5$\tabularnewline
\cline{2-3} 
 & FF and FT and FB & \begin{tabular}{c}
($*$$\uparrow$, F $>$ B $>$ T)\tabularnewline
\hline
($*$$\uparrow$, F $>$ T $>$ B)\tabularnewline
\end{tabular} & \tabularnewline
\hline
\end{tabular}
\par\end{centering}

\end{table}

We have also studied the dynamics of the TS attractors at other values
of $K$ and $(\gamma,\beta)$, but whose locations are not close to
the boundary of the TS phase. We found TS attractors with approximately
the same amplitudes of $56c_{\mathrm{o}}/25$, 2$c_{\mathrm{o}}$,
$8c_{\mathrm{o}}/5$, but we also found attractors with other amplitudes
such as $14c_{\mathrm{o}}/5$ and $72c_{\mathrm{o}}/35$. An exhaustive
search of all possible amplitudes is beyond the scope of our study.
Since they have qualitatively similar behaviors, we will continue
our analysis using only the three types of attractors we have described.

\section{Periods of the Price Cycles\label{sec:periods}}

After obtaining an estimate of the amplitudes of the price cycles
in the previous section, we can derive the periods of the price cycles
if we also know about the price change per time step. This information
is also available from Table \ref{tab:second cash redis} for type
I attractor.

Consider the falling trend from the peak price of 
$P_{\mathrm{max}}\approx56c_{\mathrm{o}}/25$
to the valley $P\approx-56c_{\mathrm{o}}/25$. 
From Tables \ref{tab:first cash redis} 
and \ref{tab:second cash redis},
we find that $A\mathrm{_{sell}}=3N/9$ from $P=56c_{\mathrm{o}}/25$
to $P=16c_{\mathrm{o}}/25$, 
and $A\mathrm{_{sell}}=5N/9$ from $P=16c_{\mathrm{o}}/25$
to $P=-56c_{\mathrm{o}}/25$. Hence, the period of the price cycles
is given by
\begin{eqnarray}
	\frac{T_{\mathrm{I}}}{2}
	&=&\left(\frac{3N}{9}\right)^{-\gamma}
	\left(\frac{56}{25}c_{\mathrm{o}}-\frac{16}{25}c_{\mathrm{o}}\right)
	\nonumber \\
  & & +\left(\frac{5N}{9}\right)^{-\gamma}\left(\frac{16}{25}c_{\mathrm{o}}
	+\frac{56}{25}c_{\mathrm{o}}\right).\label{eq:rank 1 half period}
\end{eqnarray}
This reveals the dependence of the period on $c_{\mathrm{o}}$ and
$\gamma$
\begin{equation}
	T_{\mathrm{I}}=\left(\frac{16c_{\mathrm{o}}}{25N^{\gamma}}\right)
	\left[5\left(3^{\gamma}\right)+9\left(\frac{9}{5}\right)^{\gamma}
	\right].
\label{eq:period I}\end{equation}
The periods of types II and III attractors are derived in Appendix.
To compare with simulation results, we calculate the harmonic mean
of the periods averaged over the initial conditions. 
From Table \ref{tab:3 types attractors},
the probabilities of occurrence are $1/3$, $1/6$ and $1/2$ for
types I, II and III attractors respectively. Hence the average period
of TS attractors is
\begin{equation}
	T_{\mathrm{av}}
	=\left(\frac{1}{3}T_{\mathrm{I}}^{-1}+\frac{1}{6}T_{\mathrm{II}}^{-1}
	+\frac{1}{2}T_{\mathrm{III}}^{-1}\right)^{-1}.
\label{eq:average period}\end{equation}
Substituting Eqs. (\ref{eq:period I}), (\ref{eq:per2}) and (\ref{eq:per3}),
we obtain\begin{eqnarray}
	T_{\mathrm{av}}&=&\frac{16c_{\mathrm{o}}}{N^{\gamma}}\times\nonumber \\
 & & \left\{ \frac{25}{3\left[5\left(3^{\gamma}\right)
	+9\left(9/5\right)^{\gamma}\right]}
	+\frac{1}{3\left(3^{\gamma}\right)}
	+\frac{5}{4}\left(\frac{5}{9}\right)^{\gamma}\right\} ^{-1}.
	\nonumber \\
\label{eq:final average period}\end{eqnarray}
So far we have derived this expression for a particular value of $\gamma$
($=0.2$). Observing that the attractor structures in a broad range
of $\gamma$ are qualitatively similar, we extrapolate this result
to general values of $\gamma$. Interpolating the expression by an
exponential function of $\gamma$ between $\gamma=0$ and $\gamma=1$,
we have\begin{equation}
\mathrm{ln}T_{\mathrm{av}}\approx B-m\gamma,\label{eq:lnT}\end{equation}
where $B=\mathrm{ln}\left(448c_{\mathrm{o}}/61\right)$ and 
$m=\mathrm{ln}\left(3514N/7137\right)$.

\begin{figure}
\noindent \begin{centering}
\includegraphics[width=86mm]{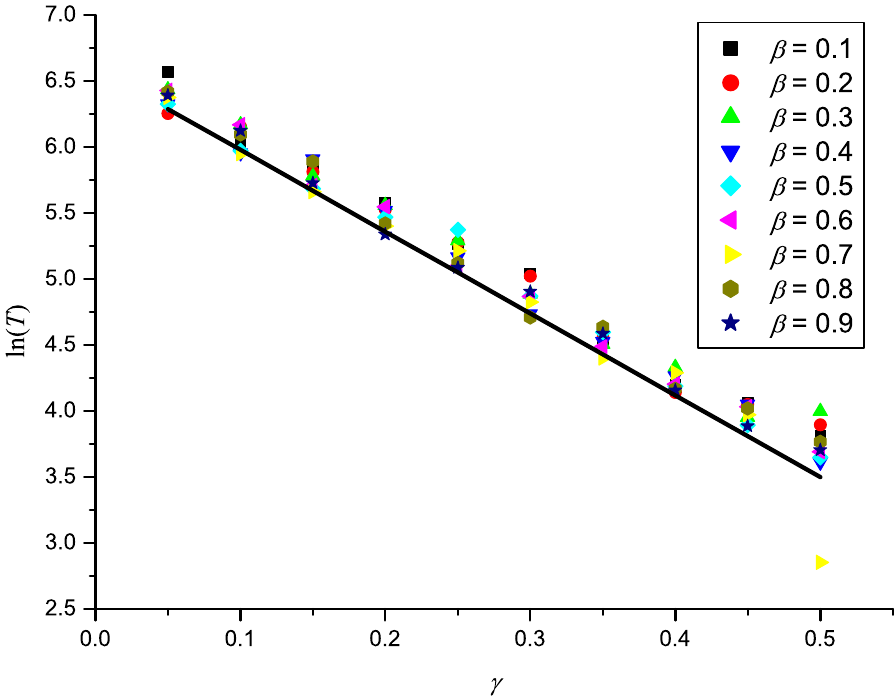}
\par\end{centering}

\caption{(Color online) The dependence of the TS period on $(\gamma,\beta)$ 
averaged over 10 samples. The line is the prediction of Eq. \ref{eq:lnT}.
The parameters are $m=2$, $s=2$, $K=3$, $N=1000$, $c_{\mathrm{o}}=100$.
\label{fig:lnT and Gamma}}

\end{figure}

Figure \ref{fig:lnT and Gamma} 
shows the average TS period for $c_{\mathrm{o}}=100$
obtained from simulations. It shows that the period is an exponential
function of $\gamma$. $B$ and $m$ in the figure are determined
experimentally to be 6.71 and 6.13 respectively. This compares favorably
with the theoretical prediction of Eq. (\ref{eq:lnT}) which yields
$B=6.60$ and $m=6.20$.

\section{The Phase Boundary\label{sec:The-Phase-Boundary}}

\begin{figure}
\noindent \begin{centering}
\includegraphics[width=86mm]{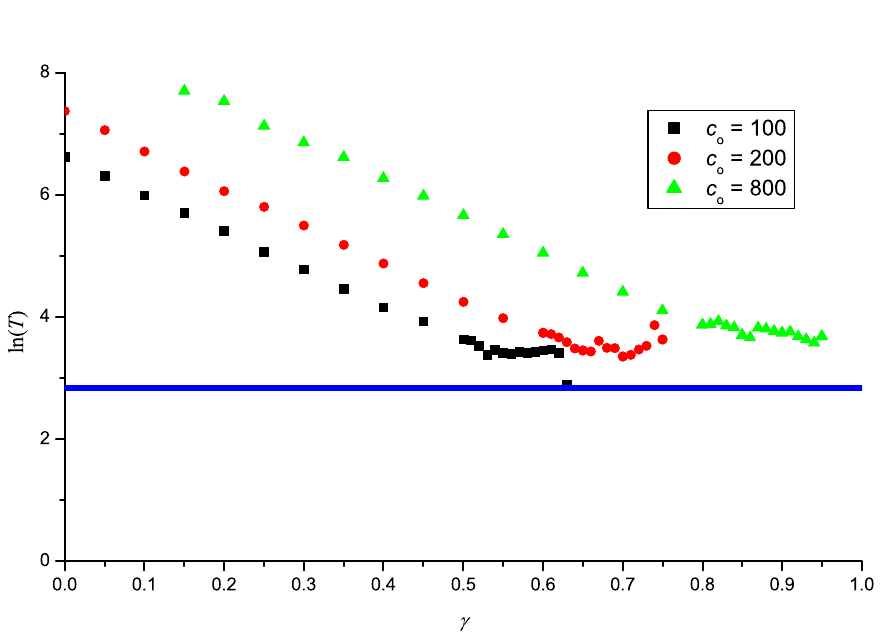}
\par\end{centering}

\caption{(Color online) The dependence of the TS period on $\gamma$ averaged over the 12
possible attractors for various $c_{\mathrm{o}}$ down to the boundary
of the TS phase. The line corresponds to the bound of $4K+5$. The
parameters are $s=2$, $K=3$, $\beta=0.5$, $N=1000$. \label{fig:period bound}}

\end{figure}

To search for the existence condition of the TS phase, we plot the
period of the TS attractors as a function of $\gamma$ for various
$c_{\mathrm{o}}$ in Fig. \ref{fig:period bound}. Remarkably, the
TS attractors disappear when the period falls below an apparently
universal value, suggesting that there is a lower bound of the TS
period around $4K+5$.

\begin{figure}
\noindent \begin{centering}
\includegraphics[width=86mm]{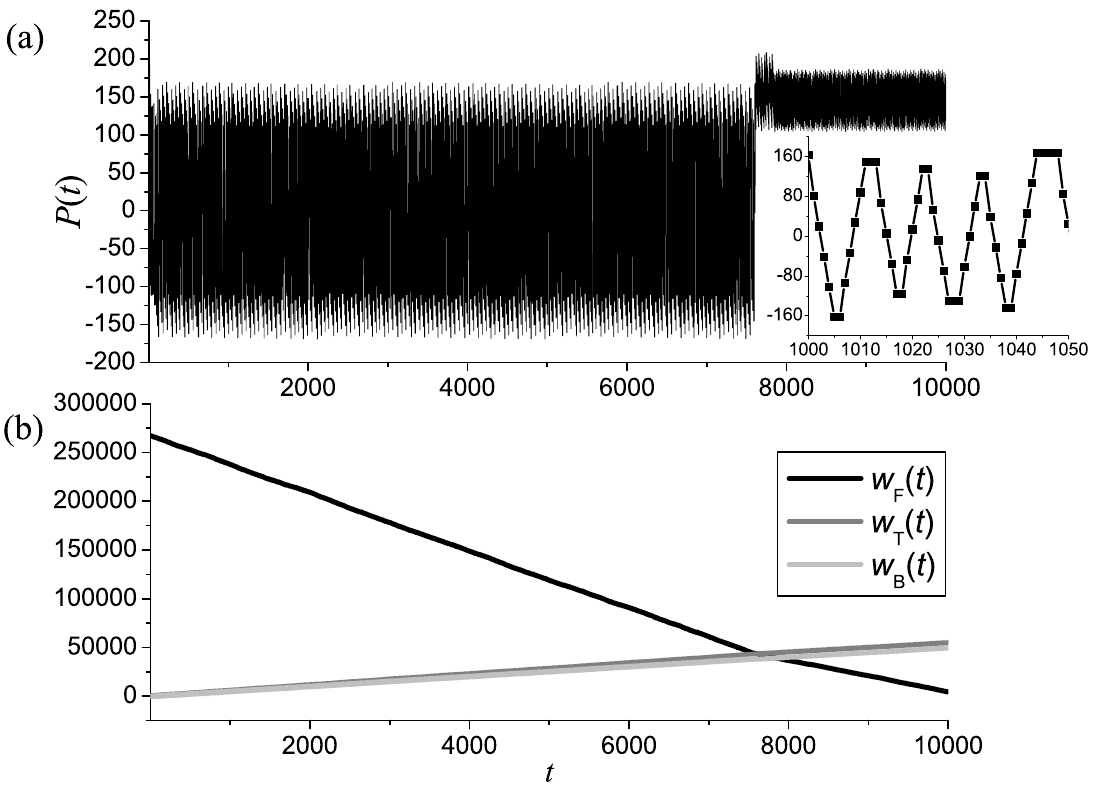}
\par\end{centering}

\caption{The transient existence of the TS attractor at 
$(\gamma,\beta)=(0.65,0.5)$
starting with the following virtual cash of the strategies: 
$c_{\mathrm{F}}=2000N^{\gamma},c_{\mathrm{T}}=c_{\mathrm{B}}=0$,
for (a) the price series, where the inset shows the magnified plot
of the TS regime and (b) the virtual wealth series. Other parameters
are: $s=2$, $K=3$, $N=1000$,$c_{\mathrm{o}}=100$. \label{fig:BO phase bias F}}

\end{figure}

This possibility is further supported by the numerical experiment
described in Fig. \ref{fig:BO phase bias F}. We start the Wealth
Game with an initial condition biased towards the F strategy, but
at a very high value of $\gamma$ deep in the BO phase. This favors
the TS attractor in the transient stage, which is expected to be destabilized
at the steady state. We observe that the virtual wealth of the F strategy
decreases from one period to another and is accompanied by periods
of the TS attractor shorter than $4K+5$. Meanwhile, the virtual wealth
of the B (or T) strategy keeps on increasing period after period.
Eventually, the F strategy is outperformed by the B strategy and the
TS attractor disappears. 

\begin{figure}
\noindent \begin{centering}
\includegraphics[width=86mm]{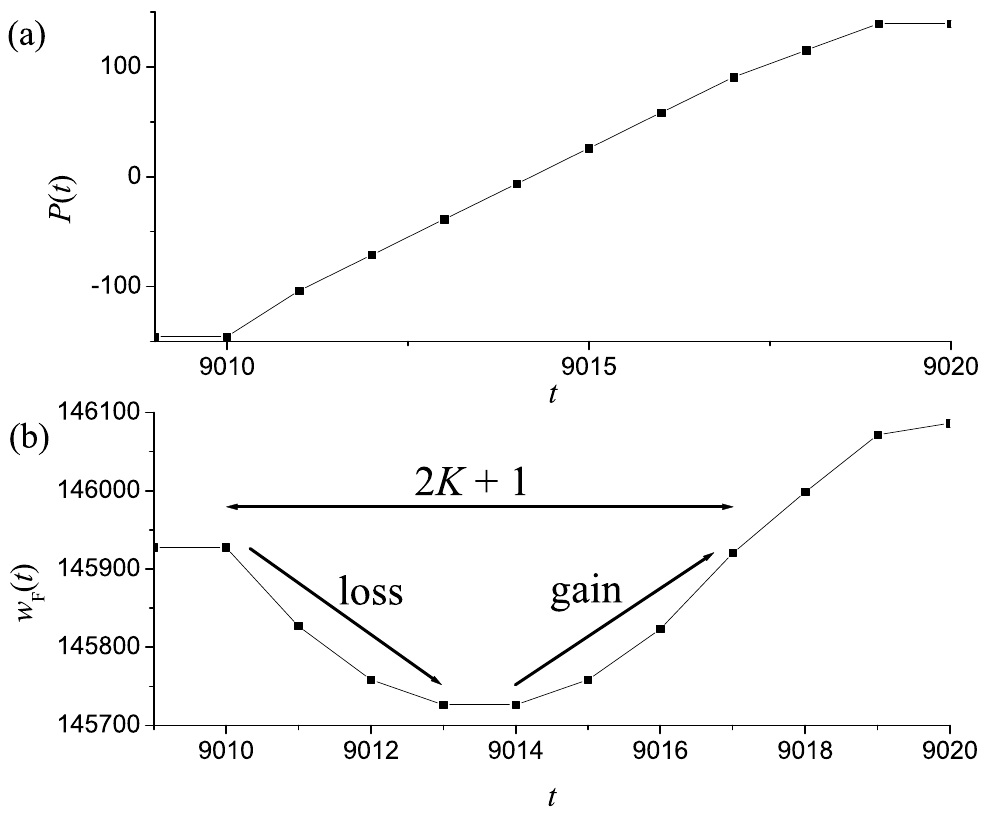}
\par\end{centering}

\caption{The time series of (a) the price and (b) the virtual wealth of the
F strategy in a half-period 
of the TS attractor at $(\gamma,\beta)=(0.55,0.8)$, $s=2$,
$K=3$, $N=1000$, $c_{\mathrm{o}}=100$. Note that in general, the
F strategy requires $2K$ steps to regain
the virtual wealth at the previous quiet period when $\beta\le 0.5$, 
and $2K+1$ steps when $\beta>0.5$.
\label{fig:4K+5}}

\end{figure}

Let us analyze the attractors with the shortest possible TS period. 
As shown in Fig. \ref{fig:4K+5} for
a TS attractor, the F strategy starts from the minimum position $-K$
at the beginning of stages 1 and 2, and takes buying actions in response
to the signal $\uparrow\uparrow$ for $2K$ steps. This continues
until it reaches the maximum position $K$. A quiet period follows.
The signal at the first step of the quiet period is $\uparrow\uparrow$,
but the signals in the following steps are random. If the signal is
$\uparrow$, the quiet period continues, but if the signal is $\downarrow$,
all strategies respond to $\uparrow\downarrow$ by selling, and the
dynamics enters stage 3. Hence the average length of the quiet period
is $1+\sum_{n=0}^{\infty}n/2^{n+1}=2$.

When the falling price trend of stage 3 starts, the F strategy makes
the right move of selling, but its position remains positive due to
the rising trend in stages 1 and 2. It takes the strategy $K$ steps
to change its position to $0$. As shown in Fig. \ref{fig:4K+5},
it is losing virtual wealth during this period of time. It takes another
$K$ steps to change its position from zero to minimum, and the strategy
is regaining virtual wealth during this period. For these $2K$ steps,
its virtual wealth gain and loss are roughly balanced. This completes
the adjustment stage of the F strategy. If the falling trend is longer
than $2K$ steps, then the strategy can start to gain virtual wealth
using its trend-following outputs and stabilize the TS attractor.

To make a more quantitative estimate, we consider the case $\gamma=0$,
in which the price change at each step is $A^{0}=1$, independent of
the volume of buying and selling quotations. By tracing the virtual
wealth change of the F strategy during a price cycle of length $4K+4$
(with an average length of 2 during the quiet periods at the peaks
and valleys), the virtual wealth gain of the F strategy is calculated
to be $2K(1-2\beta)$, whereas those of the T and B strategies are
1. When the period of the price cycle lengthens, the virtual wealth
gain of the F strategy increases by $K$ for every extension of the
period by one step. If we consider the mid-position of the phase diagram
with $\beta=1/2$, then for price cycles of period $4K+5$, the F
strategy outperforms the T and B strategies by $2K(1-2\beta)+K-1=K-1$.
Hence for $K>1$, it is reasonable to estimate the phase boundary by the condition
that the period of the price cycle becomes $4K+5\equiv T_{\mathrm{c}}$.

\begin{figure}
\noindent \begin{centering}
\includegraphics[width=86mm]{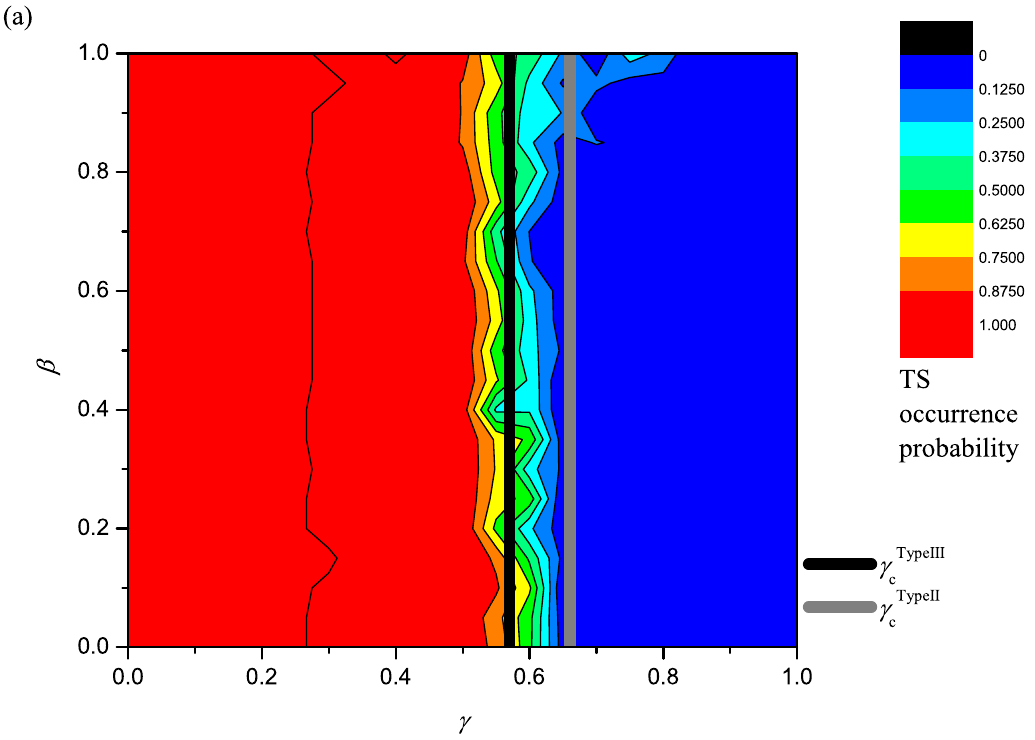}
\par\end{centering}

\noindent \begin{centering}
\includegraphics[width=86mm]{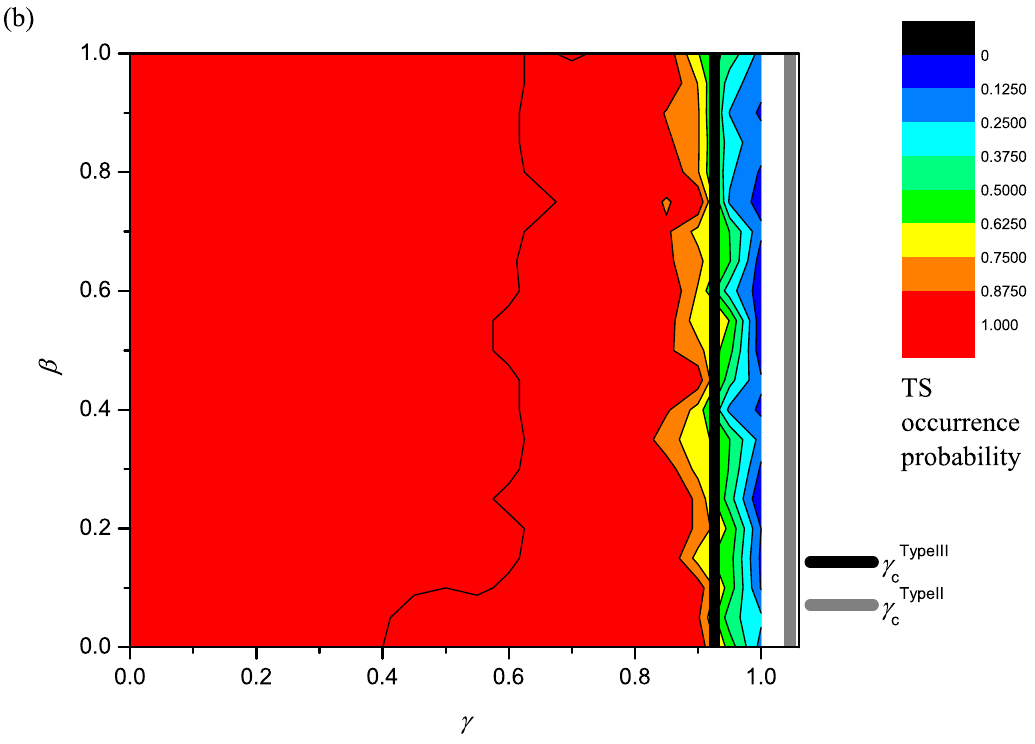}
\par\end{centering}

\caption{(Color online) Comparison of the theoretical prediction of $\gamma_{\mathrm{c}}$
with the simulation results for (a) $c_{\mathrm{o}}=100$ and (b)
$c_{\mathrm{o}}=925$. Other parameters are $s=2$, $K=3$, $N=1000$ (20 samples).
\label{fig:compare theory}}

\end{figure}

We are now ready to calculate the critical value $\gamma_{\mathrm{c}}$
of the TS attractor at the phase boundary. However, we observe in
Fig. \ref{fig:phase diagram1} that the transition to the BO phase
occurs within a narrow but finite range of $\gamma$, instead of having
an abrupt change. This is due to the dependence on the initial conditions,
as classified according to the three types of attractors. By equating
$T_{\mathrm{c}}$ to the periods in Eqs. (\ref{eq:period I}), (\ref{eq:per2})
and (\ref{eq:per3}), we find that $\gamma_{\mathrm{c}}$ takes the
values of 0.65, 0.66 and 0.57 for types I, II and III attractors respectively.
The phase lines for types II and III attractors are located in the
phase diagrams in Fig. \ref{fig:compare theory}, and the matching
with the simulation results is quite well.

\section{Dependence on Population Composition\label{sec:f-dependence}}

\begin{figure}
\noindent \begin{centering}
\includegraphics[width=86mm]{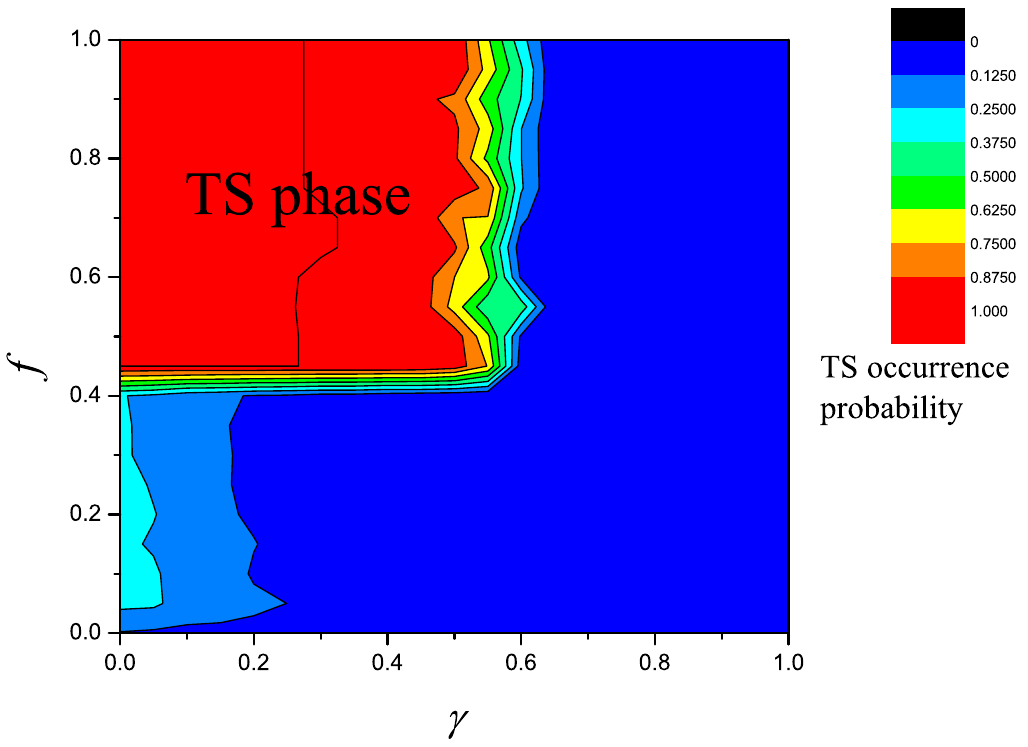}
\par\end{centering}

\caption{(Color online) The phase diagram in the space of the F strategy factor $f$ and price
sensitivity $\gamma$. Other parameters are $s=2$, $K=3$, $N=1000$,
$c_{\mathrm{o}}=100$, $\beta=0.5$ (20 samples).\label{fig:phase diagram with f}}

\end{figure}

So far, we have considered the case of unbiased assignment of strategies,
that is , $f_{\mathrm{F}}:f_{\mathrm{T}}:f_{\mathrm{B}}=1:1:1$, where
$f_{\mathrm{\sigma}}$ is the ratio of the probabilities 
of assigning strategies $\mathrm{\sigma}$
to an agent. It is interesting to consider how the market dynamics
changes when we vary the ratio of strategies 
to $f_{\mathrm{F}}:f_{\mathrm{T}}:f_{\mathrm{B}}=f:1:1$,
where $f$ is referred to as the {\it trendsetter (TS) factor}. 
Figure \ref{fig:phase diagram with f}
shows the phase diagram in the space of $f$ and $\gamma$. The TS
phase exists for sufficiently large $f$ and sufficiently small $\gamma$.
This suggests that in order to trigger the TS price trends, the market
should be dominated by enough trendsetting strategies. In other words,
a market is trendy only when there are enough trend-believing agents.
In contrast, if the game is full of T and B strategies, it becomes
a market with fundamentalists as the majority, and price trends cannot
be set up easily. Such a picture shows qualitative consistency with
the results obtained in \cite{Giardina03}, where a polarization parameter
determines the statistical weight of trend-following strategies, and
the periodic phase exists at high polarization.

\begin{figure}
\noindent \begin{centering}
\includegraphics[width=86mm]{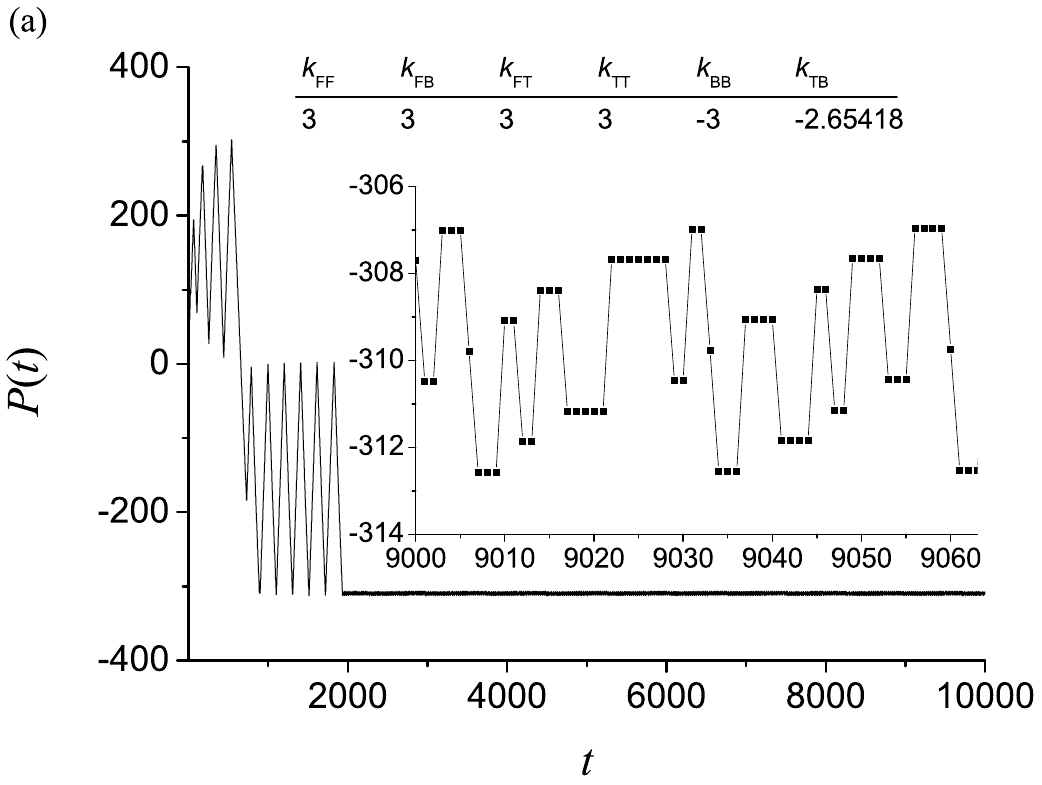}
\par\end{centering}

\noindent \begin{centering}
\includegraphics[width=86mm]{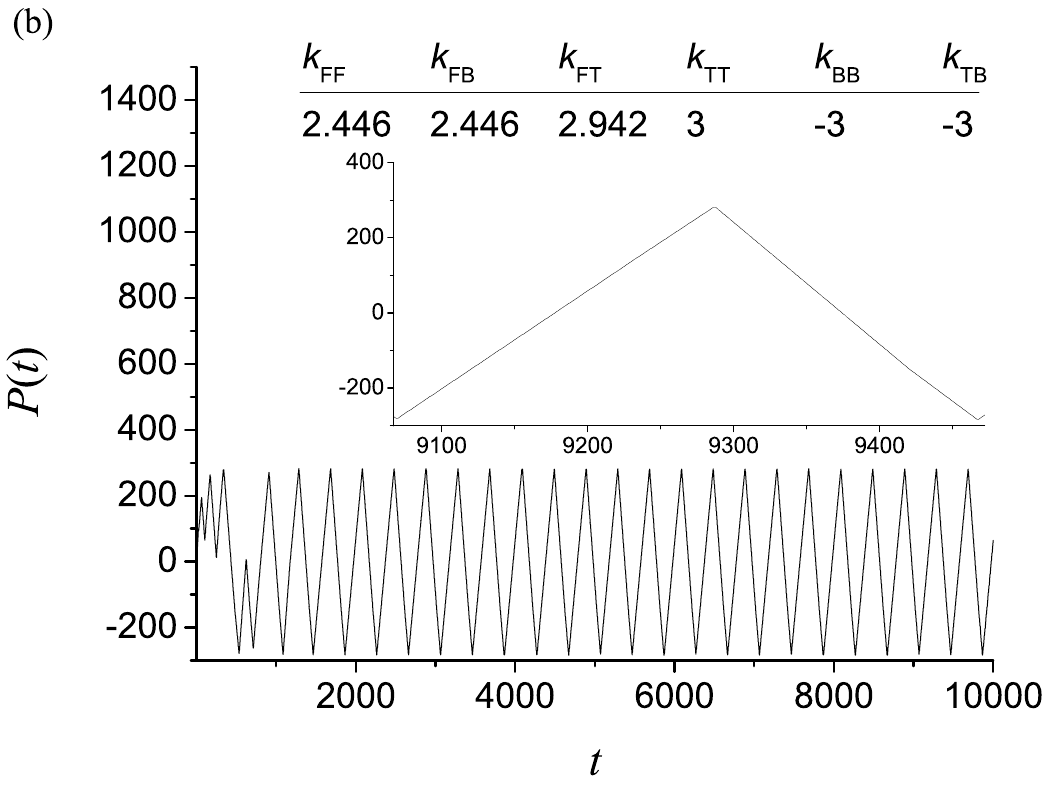}
\par\end{centering}

\caption{The price series of an attractor 
before and after undergoing the phase transition 
at (a) $f=0.4$ (b) $f=0.5$. 
The middle plot and the top table are the magnified graph 
of the price series and the positions of different agents at stable states. 
Other parameters are $m = 2$, $s = 2$, $K = 3$, $N = 1000$, 
$c_{\mathrm{o}}=100$, $(\gamma,\beta)=(0.2,0.8)$, 
initial condition $(\uparrow\uparrow,\mathrm{T} > \mathrm{F} > \mathrm{B})$.
\label{fig:phase transition in f}}

\end{figure}

\begin{figure}
\noindent \begin{centering}
\includegraphics[width=86mm]{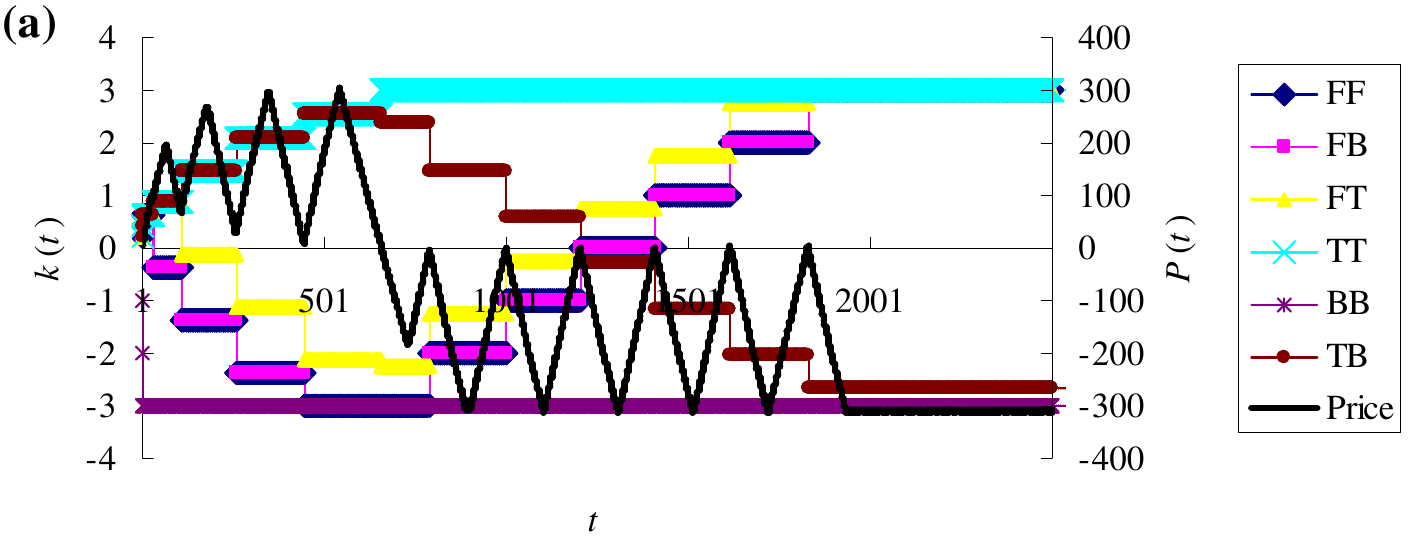}
\par\end{centering}

\noindent \begin{centering}
\includegraphics[width=86mm]{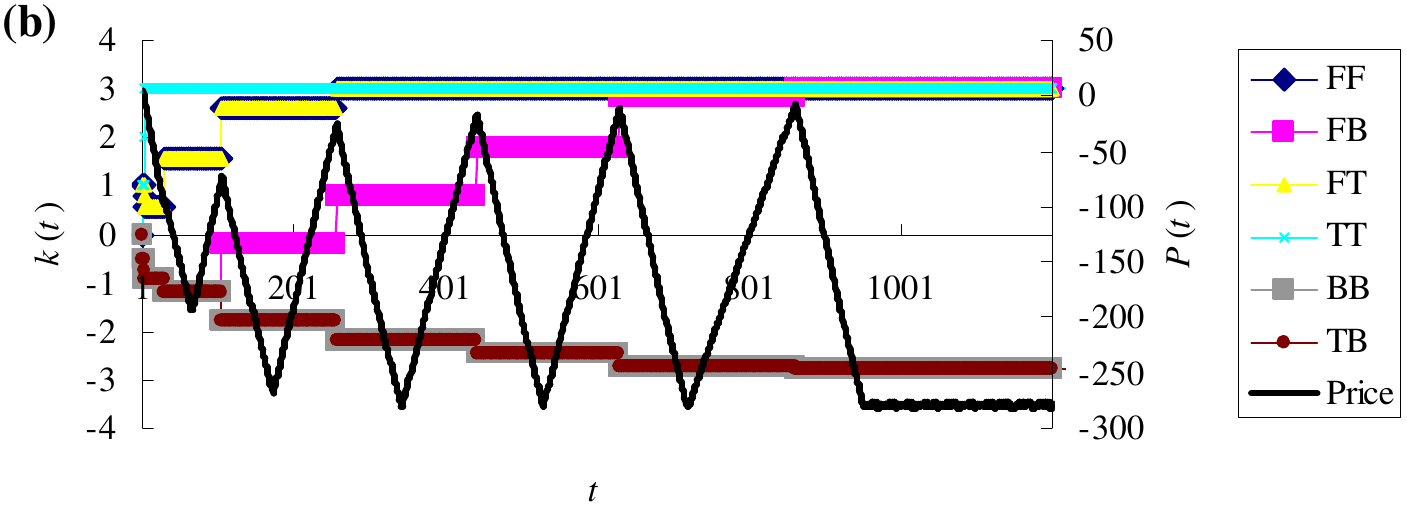}
\par\end{centering}

\caption{(Color online) Evolvement of the agents' positions for initial condition 
(a) ($\uparrow\uparrow,\mathrm{T}>\mathrm{F}>\mathrm{B}$),
(b) ($\downarrow\uparrow,\mathrm{B}>\mathrm{F}>\mathrm{T}$). The
price series is also shown for reference. Parameters: $s=2$, $K=3$,
$N=1000$, $c_{\mathrm{o}}=100$, $(\gamma,\beta)=(0.2,0.8)$, $f=0.4$.
\label{fig:f04 pos}}

\end{figure}

The vertical segment of the phase boundary 
in Fig. \ref{fig:phase diagram with f}
shows little dependence on $f$, indicating that our analysis at $f=1$
is a good approximation. 
To analyze the horizontal segment, 
we consider the typical attractor profile below and above the
phase boundary in Fig. \ref{fig:phase transition in f}.
Note that at the transient stage, the price series in both cases appear
as TS. The difference lies in the behavior on approaching the steady
state. When $f$ is small, the F-group agents (FF, FT and FB) are
the minority. In the absence of market makers, the need to balance
the buying and selling volumes has significant consequences in determining
the types of agents whose positions are {\it saturated} (that is,
reach $\pm K$). Hence, we show the evolvement of the agents' positions
in Fig. \ref{fig:f04 pos} during the separation stage for two typical
cases. In these cases the position of the F-group agents saturate
at $K$; the cases of $-K$ have the same behavior after gauge transformation.

The outcome of the separation stage is that the TT and F-group agents
take up positive positions at the steady state, and the BB and TB
agents have negative positions. Note that at the steady state, the
only active sellers with unsaturated positions are the TB agents in
Fig. \ref{fig:f04 pos}(a), and the TB and BB agents in Fig. \ref{fig:f04 pos}(b).
More significantly, the F-group agents have saturated positive positions.
Once their positions are saturated, their buying quotations disappear.
The TS price trend stops, and the steady state enters the bouncing
attractor.

\begin{figure}
\noindent \begin{centering}
\includegraphics[width=86mm]{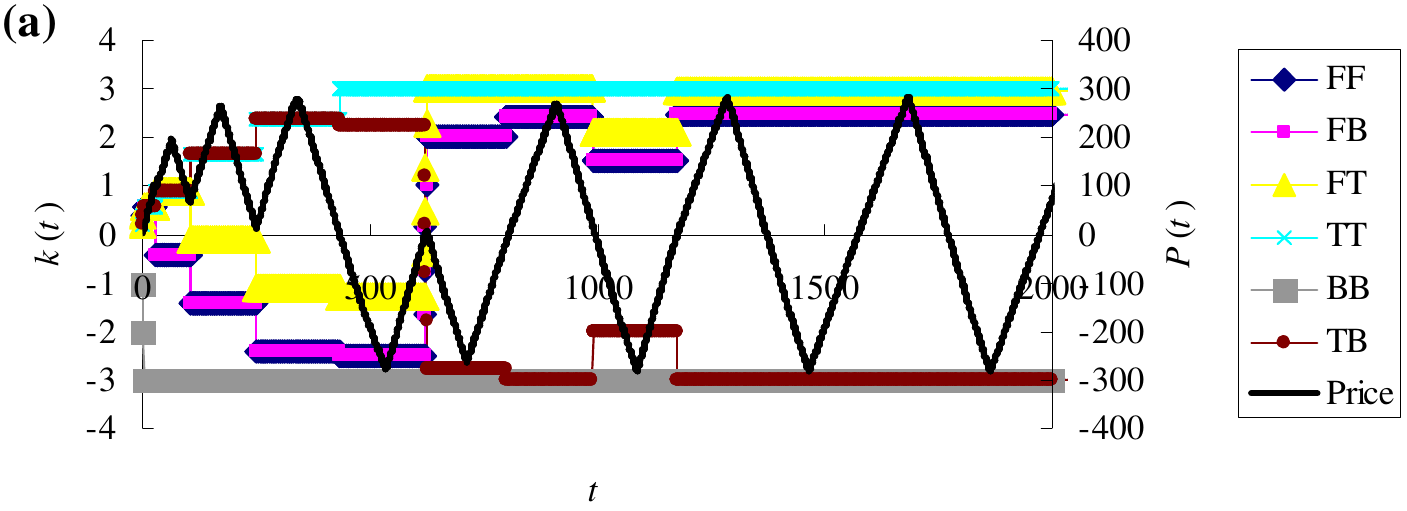}
\par\end{centering}

\noindent \begin{centering}
\includegraphics[width=86mm]{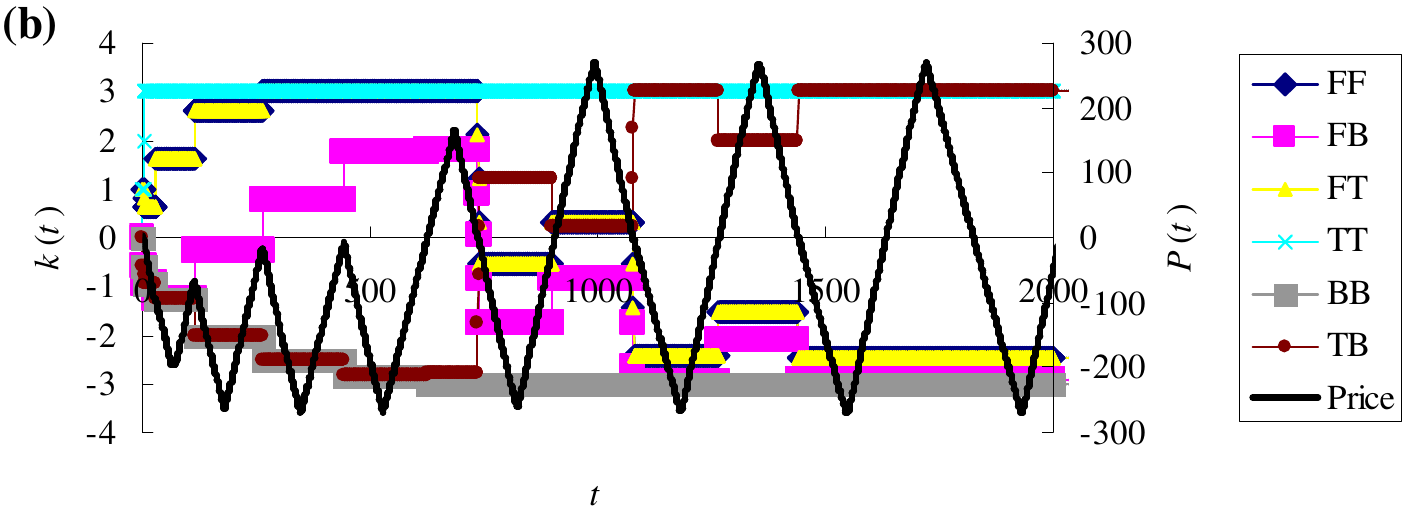}
\par\end{centering}

\caption{
(Color online) Same as Fig. \ref{fig:f04 pos}, 
except that $f=0.5$.
\label{fig:f05 pos}}

\end{figure}

This transition mechanism from the TS transient to the BO attractor
is further confirmed by the dynamics of agents' positions when $f$
is increased to enter the TS phase, as shown in Fig. \ref{fig:f05 pos}.
Since $f$ is larger when compared with Fig. \ref{fig:f04 pos},
the TB and BB agents saturate at position $-K$ before the F-group
agents reaches position $K$. The result is exactly the opposite:
the selling quotes of the TB and BB agents disappear, and the price
continues to rise. More important, the F-group agents remain unsaturated.
They emerge as active agents with the freedom to make buying and selling
quotes, and the TS attractor is sustainable at the steady state. 

This transition mechanism enables us to derive the macroscopic condition
for the disappearance of the TS phase, irrespective of the transaction
details. The total volume of stocks that can be sold by the BB and
TB agents is $V_{\mathrm{s}}=3NK/(2+f)^{2}$. The total volume of
stocks that can be bought by the TT and F-group agents is $V_{\mathrm{b}}=NK(f^{2}+4f+1)/(2+f)^{2}$.
If $V_{\mathrm{b}}<V_{\mathrm{s}}$, the F-group agents become saturated
and remain inactive. Thus we have the BO phase when $f<\sqrt{6}-2\approx0.449$.
This prediction matches well with the simulation result in Fig. \ref{fig:phase diagram with f}.

\section{\label{cha:Conclusion}Conclusion}

We have considered a simplified Wealth Game in which no market makers
are present. Since buying and selling quotations are not balanced,
the dynamics becomes complicated. Furthermore, due to the prevalence
of dry quotations, the dynamics becomes heavily dependent on the initial
conditions. This makes it difficult to analyze the model and understand
its underlying mechanism. To circumvent this difficulty, we simplify
the input dimension of the strategies to 2 and restrict the strategies
to three representative ones, namely, F, T and B. Respectively, they
represent the trend-followers/trendsetters, optimistic and pessimistic
fundamentalists in the market.

With these simplifications, we observe a dynamical transition from
the TS phase to the BO phase when $\gamma$ increases. Despite the
simplicity of the model, both phases bear characteristics of real
markets. The TS phase produces price trends that resemble bubbles
and crashes in real markets. They are observed when the agents have
enough cash, and the price movement is not very sensitive to the excess
demand. The BO phase produces price trends that are relatively steady,
with occasional up-bounces or down-bounces followed by relaxations
to fundamental prices. These price trends can also be observed in
real markets when investors have very cautious moods about the fundamental
values of the stocks. Both phases are dominated by dry quotations,
which correspond to quiet moments in real markets with very low trading
volumes. In many economic systems, such as the real estates market,
dry quotations are prevalent when the price is too high or too low.

We find that the amplitude of the price cycles is determined by the
cash level of the most liquid agents. When agents with less cash stop
their quotations, the more liquid agents can still boost up or push
down the price to higher or lower levels. Prices reach their extreme
values when the agents consider it too risky to participate. However,
the quantitative relation between the price amplitude and the initial
cash is determined by the cash redistribution process during the transient
stage. To overcome the complexity of this process, we adopt a semi-empirical
approach by analyzing the separation stage for various initial conditions,
and extrapolating the predictions to the entire TS phase. Agreement
with simulation results shows that this is a good approximation.

Our study also suggests a mechanism for the disappearance of the TS
phase. All trend-following strategies need an adaptation period when
the price trend reverses. Agents become certain of the advantages
of these strategies only when the duration of a price trend is longer
than that of the adaptation period. When the price sensitivity increases,
the price change per step increases, and the period of the price cycles
shortens. When the period becomes comparable to the adaptation period,
the TS attractor becomes unsustainable.

We also find that the composition of the population affects the market
behavior. The TS phase is present in markets where trend-following
strategies are popular. When the trend-following strategies become
less popular, we observe a phase transition to the bouncing phase.
We find that this transition is due to the fact that the positions
of the trend-followers saturate at the maximum (or minimum), and no
active agents want to buy (or sell) when they decide to sell (or buy).
This allows us to derive a macroscopic condition for the phase transition
by balancing the volume of supply and demand of stocks. The prediction
agrees with the simulation results well.

It is interesting to compare our model with the Wealth Game with the
market makers present \cite{Yeung08}. In both cases, the TS phase
exists due to the presence of trend-followers. However, when market
makers are present, the price trend is driven by real transactions
and has a slightly different dynamics. For example, the so-called
fickle agents are those who hold a T and B strategy, and fickle their
preference between the two strategies with a delayed response. They
push the price further up in a rising trend, and down in a falling
trend, thus creating opportunities for the trend-followers to gain
wealth. In our model without market makers, the fickle agents do not
play an important role. During the separation stage, their cash is
reduced to a very low level, as evident in Tables \ref{tab:second cash redis},
\ref{tab:type II cash redistribution} and 
\ref{tab:type III cash redistribution}
for types I to III attractors respectively. 
Hence they only play a minor role in the price dynamics.

While our model is successful in explaining and interpreting a number
of market phenomena, it can be further improved to address a broader
range of issues. One possible modification to this model is to implement
the injection of cash to the market. This may help to relieve the
problem of too many dry quotations in the present model, whose agents
are restricted by the cash rule. Furthermore, since the present model
is a close system with constant average wealth, incorporating cash
injection may cause the market to evolve spontaneously towards states
which have maximal attraction of capital. In this way, it will also
address the issue of the self-organization of markets, which has drawn
considerable attention in recent models 
\cite{Giardina03, Challet01, Yeung08, Alfi09a, Alfi09b}.

\section*{Acknowledgement}

We thank Jack Raymond for discussions on the public dice. 
This work is supported by the Research Grants Council of Hong Kong 
(grant nos. HKUST 630607 and 604008).

\appendix*
\section{Periods of Types II and III Attractors}

In a type II attractor, there is only one significant cash redistribution
event during the separation stage. It takes place during a falling
trend, and the order of priority of the strategies is F $>$ T $>$
B. Both FB and BB agents have reached their minimum positions. As
calculated in Table \ref{tab:type II cash redistribution}, 
$A_{\mathrm{buy}}=A_{\mathrm{sell}}=3N/9$,
and the maximum final cash is $2c_{\mathrm{o}}$. 
Hence $P_{\mathrm{max}}\approx2c_{\mathrm{o}}$.

The period of type II attractor is \begin{equation}
T_{\mathrm{II}}=\frac{8c_{\mathrm{o}}}{N^{\gamma}}3^{\gamma}.
\label{eq:per2}\end{equation}

\begin{table*}
\caption{The outputs, positions, decisions and final cash (in multiples of
$c_{\mathrm{o}}$) of the agents at the cash redistribution event
of a type II attractor.\label{tab:type II cash redistribution}}
\noindent \begin{centering}
\begin{tabular}{|c|c|c|c|c|c|c|}
\hline 
Agents & Fraction & Strategy & Output & Position & Decision & Final cash\tabularnewline
\hline
\hline 
FF & $1/9$ & F & sell & $\geq-K+1$ & sell & 2\tabularnewline
\hline 
FT & $2/9$ & F & sell & $\geq-K+1$ & sell & 2\tabularnewline
\hline 
FB & $2/9$ & F & sell & $=-K$ & hold & 1\tabularnewline
\hline 
TT & $1/9$ & T & buy & $\leq K-1$ & buy & 0\tabularnewline
\hline 
TB & $2/9$ & T & buy & $\leq K-1$ & buy & 0\tabularnewline
\hline 
BB & $1/9$ & B & sell & $=-K$ & hold & 1\tabularnewline
\hline
\end{tabular}
\par\end{centering}

\end{table*}

In a type III attractor, there is often only one significant cash
redistribution event during the separation stage. It takes place
during a falling trend, and the order of priority of the strategies
is F $>$ T $>$ B. Only the BB agents have reached their minimum
positions. As calculated in Table \ref{tab:type III cash redistribution},
$A_{\mathrm{buy}}=3N/9$ and $A_{\mathrm{sell}}=5N/9$, and the maximum
final cash is $8c_{\mathrm{o}}/5$. 
Hence $P_{\mathrm{max}}\approx8c_{\mathrm{o}}/5$.

\begin{table*}
\caption{The outputs, positions, decisions and final cash (in multiples of
$c_{\mathrm{o}}$) of the agents at the cash redistribution event
of a type III attractor.\label{tab:type III cash redistribution}}
\noindent \begin{centering}
\begin{tabular}{|c|c|c|c|c|c|c|}
\hline 
Agents & Fraction & Strategy & Output & Position & Decision & Final cash
\tabularnewline
\hline
\hline 
FF & $1/9$ & F & sell & $\geq-K+1$ & sell & $8/5$\tabularnewline
\hline 
FT & $2/9$ & F & sell & $\geq-K+1$ & sell & $8/5$\tabularnewline
\hline 
FB & $2/9$ & F & sell & $\geq-K+1$ & sell & $8/5$\tabularnewline
\hline 
TT & $1/9$ & T & buy & $\leq K-1$ & buy & 0\tabularnewline
\hline 
TB & $2/9$ & T & buy & $\leq K-1$ & buy & 0\tabularnewline
\hline 
BB & $1/9$ & B & sell & $=-K$ & hold & 1\tabularnewline
\hline
\end{tabular}
\par\end{centering}

\end{table*}

Besides the single event described 
in Table \ref{tab:type III cash redistribution},
there are also initial conditions which result in TT and TB holding
slightly different amount of cash before significant cash redistribution
occurs. This gives rise to the occurrence of two events, but the final
cash distribution remains the same.

The period of type III attractor is
\begin{equation}
T_{\mathrm{III}}=\frac{32c_{\mathrm{o}}}{5N^{\gamma}}
\left(\frac{9}{5}\right)^{\gamma}.
\label{eq:per3}\end{equation}


\end{document}